\documentclass[preprint,12pt]{elsarticle}

\usepackage{amssymb}
\usepackage{amsmath}
\usepackage{bm}
\usepackage{siunitx}
\usepackage{here}
\usepackage{comment}
\usepackage{threeparttable}[h]
\usepackage{lineno}
\usepackage{color}
\usepackage{ulem}
\usepackage{subfig}
\setcitestyle{square}

\journal{Elsevier}

\begin{document}

\begin{frontmatter}



\title{Event reconstruction of Compton telescopes\\using a multi-task neural network}

\author[1,2]{Satoshi Takashima \corref{corresponding}}
\cortext[corresponding]{Corresponding author}
\ead{satoshi.takashima@phys.s.u-tokyo.ac.jp}

\author[1,3,4]{Hirokazu Odaka}
\author[2]{Hiroki Yoneda}
\author[5]{Yuto Ichinohe}
\author[1,4]{Aya Bamba}
\author[6]{Tsuguo Aramaki}
\author[7]{Yoshiyuki Inoue}

\address[1]{Department of Physics, The University of Tokyo, 7-3-1 Hongo, Bunkyo, Tokyo 113-0033, Japan}
\address[2]{RIKEN Nishina Center, 2-1 Hirosawa, Wako, Saitama 351-0198, Japan}
\address[3]{Kavli Institute for the Physics and Mathematics of the Universe (WPI), The University of Tokyo, 5-1-5 Kashiwanoha, Kashiwa, Chiba 277-8583, Japan}
\address[4]{Research Center for the Early Universe, The University of Tokyo, 7-3-1 Hongo, Bunkyo, Tokyo 113-0033, Japan}

\address[5]{Department of Physics, Rikkyo University, 3-34-1 Nishi Ikebukuro, Toshima, Tokyo 171-8501, Japan}
\address[6]{Northeastern University, 360 Huntington Ave, Boston, MA 02115, USA}
\address[7]{Department of Earth and Space Science, Osaka University, 1-1 Machikaneyama, Toyonaka, Osaka 560-0043, Japan}

\begin{abstract}
We have developed a neural network model to perform event reconstruction of Compton telescopes. This model reconstructs events that consist of three or more interactions in a detector.
It is essential for Compton telescopes to determine the time order of the gamma-ray interactions and whether the incident photon deposits all energy in a detector or it escapes from the detector. 
Our model simultaneously predicts these two essential factors using a multi-task neural network with three hidden layers of fully connected nodes.
For verification, we have conducted numerical experiments using Monte Carlo simulation, assuming a large-area Compton telescope using liquid argon to measure gamma rays with energies up to \SI{3.0}{MeV}.
The reconstruction model shows excellent performance of event reconstruction for multiple scattering events that consist of up to eight hits.
The accuracies of hit order prediction are around \SI{60}{\%} while those of escape flags are higher than \SI{70}{\%} for up to eight-hit events of $4\pi$ isotropic photons.
Compared with two other algorithms, a classical model and a physics-based probabilistic one, the present neural network method shows high performance in estimation accuracy particularly when the number of scattering is small, 3 or 4.
Since simulation data easily optimize the network model, the model can be flexibly applied to a wide variety of Compton telescopes.
\end{abstract}


\end{frontmatter}

\section{Introduction}
\label{sec:introduction}
Observation of cosmic MeV gamma rays has rich potential to illuminate various high-energy astrophysical phenomena, e.g., heavy-element nucleosynthesis\cite{Churazov2014} and cosmic-ray acceleration\cite{Bloemen1997}.
Compton telescopes are the most promising way of imaging, spectroscopy, and polarimetry of those MeV gamma rays\cite{Schoenfelder1993,Hitomi2018,Kierans2020}.
A Compton telescope detects interactions of an incident gamma ray originated from a celestial source and measures the energy deposit and the position of each hit---in this paper, a \textit{hit} is defined as a distinct chunk of energy losses within the detector material associated with a single photon interaction.
Gamma ray interactions in sub-MeV and MeV gamma-ray bands are almost occupied by photo-absorptions, Compton scatterings, and pair creations while Rayleigh scattering is usually negligible at this energy band.
Data reduction of a Compton telescope can be divided into two steps: an event reconstruction and high-level statistical analysis.
The first step reconstructs the energy and incoming direction of an incident photon that defines an \textit{event}, performed independently for a single event by using the physics of Compton scattering.
Then, the second step treats accumulated events by statistical methods to obtain information about celestial sources via imaging \cite{Strong1995,Ikeda2014} and/or spectroscopy\cite{Arnaud1996,Atwood2009}.

Events of detected gamma rays in a Compton telescope have various physical properties, usually categorized by the number of hits.
The Compton Gamma-Ray Observatory, which operated in 1991--2000, had a Compton telescope (COMPTEL) that achieved the best sensitivity in an energy band of 0.75--30 MeV. It was, however, only able to analyze events with two hits composed of the first Compton scattering and the second photo-absorption.
This limitation is one of the leading causes  of its low detection efficiency and its small effective area around 20--\SI{50}{cm^2}.
COSI\cite{Tomsick2019} was selected as a new space telescope by NASA's Explorers Program in 2021, which will promote other proposed MeV gamma-ray missions, e.g., GRAMS\cite{Aramaki2020}, AMEGO\cite{McEnery2019}, and e-ASTROGAM\cite{Angelis2017}. Those next-generation telescopes allow gamma rays to interact with detector materials more than twice.
Incorporating such multiple-hit events into the reconstruction analysis is necessary for high detection efficiency.

To estimate the energy and direction of an incident photon that causes such multiple hits, we need to determine the hit order and whether all the photon energy is deposited in a detector (or the photon escapes from the detector's active region)\cite{Kamae1987}. However, modern Compton telescopes are implemented in relatively small spaces and do not have enough hit-to-hit distances to determine the interaction order by time of flight.
In addition, conventional analysis techniques usually assume that gamma rays deposit all the energy in a detector (those events are called fully-absorbed events), and the escape events are negligible in many cases\cite{Boggs2000,Oberlack2000,Zoglauer2005}.
This can result in an underestimation of the initial energies. Note that \textit{escape} refers to the escape of a primary photon, and does not include escapes of secondary particles such as recoil electrons via Compton scattering and fluorescence X-rays, which could degrade the performance of a Compton telescope.

Thus, it is essential for high sensitivity as well as large effective areas to establish an analysis algorithm that can deal with both escape and fully-absorbed events properly.
We have already developed a physics-based probabilistic model to deal with the above complexity\cite{Yoneda2022}.
It shows excellent performance for Compton telescopes that are composed of a single dense material.
Thus, we adopt this algorithm within a standard pipeline of the Compton telescope of the GRAMS project, which is based on a large-volume liquid argon time projection chamber technology\cite{Aramaki2020}.

Another independent and promising approach is a data-driven method, namely, a neural network.
A deep neural network is capable of constructing a flexible model and is already applied to various observational techniques, such as X-ray polarimetry and gravitational wave detection\cite{Kitaguchi2019,Biswas2013,George2018}.
A neural network model with a single hidden layer was proposed for reconstructing multiple Compton scattering events\cite{Zoglauer2007a}.
However, it assumes that all events are fully-absorbed ones and calculates the initial energy by summing all the energy deposits.
This assumption can underestimate the initial energy for escape events. Thus, the initial energies of escaping photons should be estimated in other proper ways.

We, therefore, develop a multi-task neural network that performs two classifications: whether the event is a fully-absorbed event or not, in addition to the hit order.
This multi-task learning can restrain overfitting for both the tasks compared with training two independent models because such models have common layers for multiple tasks, and this structure works as inductive bias\cite{Caruana1997}.

In this paper, we present a new comprehensive method to reconstruct multiple-hit events ($n\geq 3$; $n$ is the number of hits), including escape events.
In Section \ref{sec:model}, we describe a principle to deal with multiple scattering events. Then we introduce a neural network model to determine the hit order and whether a photon escapes or not.
In Section 3, a configuration of numerical experiments is explained for performance examination of the model.
In Section 4, we demonstrate the capability of the model through the results of the numerical experiments.
In section 5, we discuss our results and potentials of neural networks, and we give our conclusions in Section 6.

\section{Model}
\label{sec:model}
At the beginning of this section, we review the principle of Compton telescopes. The essential part of the event reconstruction of an event obtained with a Compton telescope is to predict the hit order and the escape flag, which is a binary flag that indicates whether the event is a fully-absorbed event or not. Then, we describe a neural network model that predicts these two factors.

\subsection{Principle of event reconstruction}
\label{sec:model_1}
\begin{figure}[htbp]
    \centering
    \includegraphics[width=5cm]{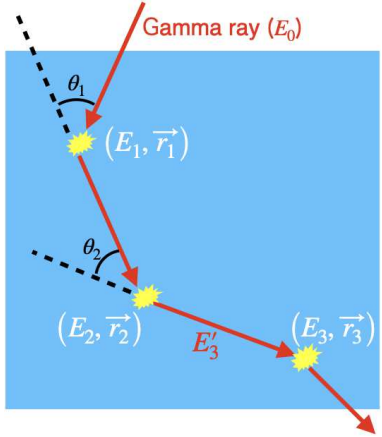}
    \caption{A schematic of gamma-ray interaction for a three-hit escape event. Detector material is painted in blue in this figure.}
    \label{fig:schematic_gamma}
\end{figure}

We consider an event that is composed of $n$ hits ($n\geq 3$); $n$ is the number of hits), which is called an $n$-hit event.
A pair of $i$-th energy deposit and position is denoted by $(E_i, \vec{r}_i)$, as shown in Figure \ref{fig:schematic_gamma}.
Position $\vec{r}_i$ is a vector that has three components $(x_i,y_i,z_i)$ in a Cartesian coordinate system.
The initial gamma-ray energy $E_0$ of a fully-absorbed event is estimated as the sum of all energy deposits.
In the case of an escape event, estimation of the initial energy $E_0$ utilizes the second scattering angle $\theta_2$\cite{Kurfess2000}, which is determined geometrically as
\begin{equation}
    \label{eq:cos_geometry}
    \cos \theta_2 = \frac{\vec{r}_{2}-\vec{r}_{1}}{|\vec{r}_{2}-\vec{r}_{1}|} \cdot \frac{\vec{r}_{3}-\vec{r}_{2}}{|\vec{r}_{3}-\vec{r}_{2}|}\ ,
\end{equation}
and the gamma-ray energy after the second scattering $E'_3$, which can be calculated by
\begin{equation}
    \label{eq:corrected_energy}
    E'_{3} = -\frac{E_2}{2} + \sqrt{\frac{E^2_2}{4} + \frac{m_\mathrm{e} c^2 E_2}{1-\cos{\theta_2}}}\ ,
\end{equation}
where $c$ and $m_\mathrm{e}$ are the speed of light and the electron mass, respectively.
Once $E'_3$ is determined, the initial energy $E_0$ is estimated by
\begin{equation}
    \label{eq:estimated_energy_escape}
    E_0 = E_1+E_2+E'_3\ .
\end{equation}
With the energy-momentum conservation law, the first scattering angle $\theta_1$ is given by
\begin{equation}
    \label{eq:compton}
    \theta_1 = \arccos{\left(1 - m_\mathrm{e} c^2 \left(\frac{1}{E_0 - E_1} - \frac{1}{E_0}\right)\right)}\ .
\end{equation}
Thus, the position of the gamma-ray source is constrained to a so-called Compton cone, which is a conic surface defined by the main axis parallel to $\vec{r}_1 - \vec{r}_2$ and the opening angle $\theta_1$.
In brief, it is essential to determine the hit order and the escape flag of an event to constrain the gamma-ray source position to a Compton cone.
In the hit order prediction, the determination of the first three hits is sufficient for the event reconstruction.

\subsection{A multi-task neural network model}
\label{sec:model_2}

We construct a neural network model for the event reconstruction.
This model is designed to predict the correct order of the first three hits and the escape flag from the measured energy deposits and positions of all hits of an event.
For convenience, we arrange the measured hits in the ascending order of the energy deposits as
\begin{gather}
    \label{eq:energy_ineq}
    \{(E,\vec{r})\} \equiv \left\{\left(E_1,\vec{r_1}\right),\left(E_2,\vec{r_2}\right),\ldots ,\left(E_n,\vec{r_n}\right)\right\}\ ,\\
    E_1 \leq E_2 \leq \ldots \leq E_n\ .
\end{gather}
Note that the subscripts $i$ of energy deposits and positions no longer indicate the time order.
This sorting works as a preliminary step by a simple physical consideration of Compton scattering before a full simulation-based machine learning.
Since it is sufficient to reconstruct both fully-absorbed and escape events, our model identifies hits from the first to the third.
Furthermore, the number of hit order candidates (permutations) is $P(n,3)=n!/(n-3)!$.
We construct a fully connected multi-task neural network to predict the scattering order and the escape flag simultaneously.  

To describe the network, we introduce a hit order classification variable $t_{k,u}$ of $k$-th event to indicate one of the $P(n,\,3)$ patterns, where $u$ is a class label index running from 1 to $P(n,\,3)$.
When the class label is $m$ among the $P(n,\,3)$ patterns, this variable is expressed by
\begin{equation}
    t_{k,u} = \begin{cases}
        1 & (u=m) \\
        0 & (u\neq m)
    \end{cases}.
\end{equation}
Then, we define a binary variable $l_k$ as an escape flag of $k$-th event:
\begin{equation}
    l_{k} = \begin{cases}
        0 & (\mathrm{fully absorbed}) \\
        1 & (\mathrm{escape})
        \end{cases}.
\end{equation}
The output of the multi-task model is a pair of two probabilities.
The first one, $\hat{y}_{k,u}$, denotes a probability that the hit order label of the $k$-th event is $u$.
The second, $\hat{z_k}$, is a probability that an event is an escape event.
The loss function of the model must be constructed to predict both the hit orders and the escape flags.
Therefore, the loss function $L_{\mathrm{CB}}$ is the sum of cross-entropy and binary cross-entropy:
\begin{align}
    \label{eq:loss}
    L_{\mathrm{CB}}\left(\left\{t\right\},\left\{l\right\},\left\{\hat{y}\right\},\left\{\hat{z}\right\}\right)&= -\frac{1}{N}\sum_{k:\mathrm{event}}^N \biggl[l_k\ln{\hat{z}_k} + (1-l_k)\ln{\left(1-\hat{z}_k\right)} \nonumber\\ 
    &+ \sum_{u=1}^{P(n,\,3)}t_{k,u} \ln{\hat{y}_{k,u}}\biggr]\ ,
\end{align}
where $N$ is a batch size in one update of the model parameters.
In summary, the model converts an input hit array $\{(E_i, x_i, y_i, z_i)\}$ into an output, $(t_{k,u},l_k)$, composed of the two probabilities for the hit order label and the escape flag.
We finally choose a candidate whose probabilities of the hit order and the escape flag are the highest.

\begin{figure}[htbp]
    \centering
    \includegraphics[width=5cm]{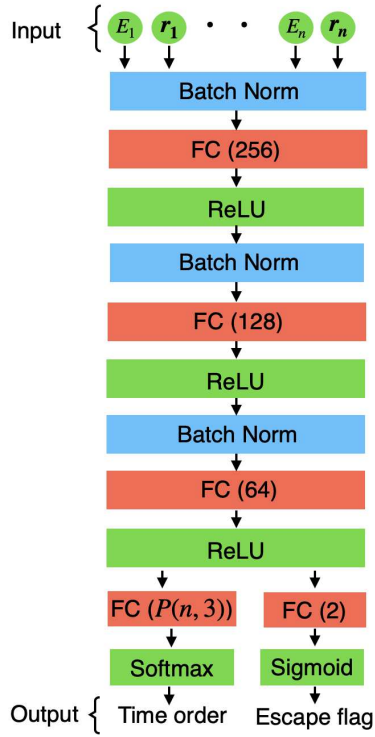}
    \caption{The network model of the event reconstruction. Three fully connected layers (FCs) are stacked with activation functions and batch normalization layers. The number of each node of fully connected layers is written in a parenthesis after FC.}
    \label{fig:DualLabelNN}
\end{figure}

Figure \ref{fig:DualLabelNN} shows the architecture of the network.
The network includes three fully connected layers, and each of them includes a batch normalization layer for suppressing overfitting\cite{Ioffe2015}.
An activation function on the hit order prediction applies a softmax function, which is widely used in multiclass classification.
On the other hand, a sigmoid function is used to produce a probability on the escape flag.

In summary, to reconstruct an $n$-hit event all the energy deposits $\{E_i\}_{i=1,\ldots,n}$ and coordinate components of positions $\{(x_i,y_i,z_i)\}_{i=1,\ldots,n}$ are input into the neural network model. The dimension of the input data for every event is $4n$. The model predicts two labels: a hit order and an escape flag. It should be stressed that difference between our method and previous work that utilizes a neural network\cite{Zoglauer2007a} is not only the existence of escape flags but also the dimensions of input data. The dimension of input data gets explosively large as $n$ increases, since the previous model takes as inputs many kinds of information about the physics of gamma-ray interaction for each sequence permutation. Therefore, that model needs even larger computer memory (if $n=6$, the dimension of the inputs becomes 20184). On the other hand, the dimension of input data in our model increases linearly. The dimension of input data is 32 even if $n=8$, which demonstrates that our model can be implemented with smaller computer resources.

\section{Numerical experiments}
\label{sec:numerical_experiments}
In this section, we conducted numerical experiments using Monte Carlo simulation to evaluate the performance of the developed neural network model. 
The model is implemented with Google Tensorflow 2.3 \cite{Mart2015} with CPU, Intel Core i9-9960X and graphics processing units (GPU), NVIDIA TITAN RTX.

\subsection{Simulation configuration}
\label{sec:numerical_experiments_1}
For the network verification and validation, we use a simulation model based on the GRAMS project. GRAMS utilizes a liquid argon time projection chamber (LArTPC)\cite{Rubbia1977} as a Compton telescope.
This detector is suitable to test the model since multiple Compton scatterings efficiently occur in that large-volume detector.
We simulated MeV gamma-ray events using ComptonSoft\cite{Odaka2010}, which is a framework that has been developed mainly for simulation and data analysis of Compton telescopes, and performed Monte Carlo simulation of gamma rays based on the Geant4 toolkit\cite{Agostinelli2003,Allison2006,Allison2016} through ComptonSoft.
We set a cuboid of \SI{140}{cm}\,$\times$\,\SI{140}{cm}\,$\times$\,\SI{20}{cm} filled with liquid argon, which is the same configuration of the main active volume of the GRAMS detector\cite{Aramaki2020}.

In the simulations in this work, the detector was assumed to be a two-dimensional pixel detector, and each pixel has depth sensitivity.
The energy resolution $\sigma_E$ (standard deviation) was set to
\begin{equation}
    \sigma^2_E = \left(0.5 \sqrt{E_{\mathrm{keV}}}\right)^2 + (5\,\mathrm{keV})^2
\end{equation}
where $E_{\mathrm{keV}}$ is an energy deposit in units of \si{keV}. 
The energy threshold of the detector was set to be \SI{25}{keV}.
The pixel size was \SI{2}{mm}, which determined the spatial resolution for the $xy$ directions, and the detected $z$-position was smeared by a Gaussian function whose standard deviation is \SI{1}{mm}.
When energy is deposited at two adjacent pixels, their signals were merged into one with an energy of the sum of these two.

\subsection{Angular resolution measure}
\label{sec:numerical_experiments_2}
The angular resolution measure (ARM) is often used for performance evaluation of a Compton telescope.
If the hit order and the initial gamma-ray energy of an event are determined, the first kinematic scattering angle $\theta_{\mathrm{K}}$ is calculated using Equation (\ref{eq:compton}).
If the source direction is known, the first scattering angle $\theta_{\mathrm{G}}$ is also given by the position measurement.
The ARM is then defined as the difference between these two scattering angles:
\begin{equation}
    \mathrm{ARM} \equiv \theta_{\mathrm{K}} - \theta_{\mathrm{G}}\ .
\end{equation}
A sharp peak around $\mathrm{ARM}=0$ means a good angular resolution and that a systematic offset is small.
In addition to the angular resolution, we evaluate the distribution of reconstructed energy when monoenergetic gamma rays are observed.

\subsection{Data processing for training and evaluation}
\label{sec:numerical_experiments_3}
\begin{figure}[htbp]
    \centering
    \includegraphics[width=6cm]{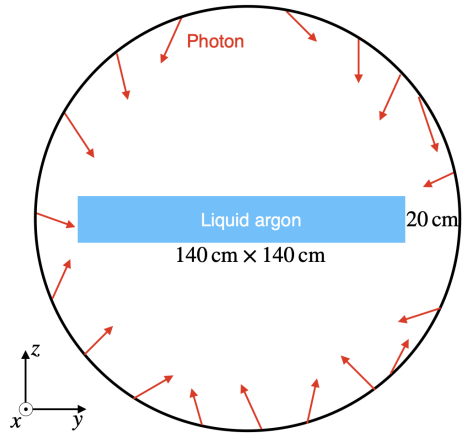}
    \caption{Configuration of the Monte Carlo simulation. The detector in the center is assumed to be the GRAMS telescope that utilizes a liquid argon time projection chamber.}
    \label{fig:isotropic}
\end{figure}

We performed training and evaluation of our network model.
This work covers $n$-hits with $3\le n \le 8$.
We suppose isotropic gamma rays from a homogeneous distribution ($4\pi\,$steradian) (see Figure \ref{fig:isotropic}).
The network was trained on generated gamma-ray events whose energies are uniformly distributed up to \SI{3}{MeV}. This band includes many interesting gamma-ray lines; \SI{0.511}{MeV} (positron annihilation), \SI{1.157}{MeV} ($^{44}\mathrm{Ti}$), and \SI{2.2}{MeV} (neutron capture). An energy band higher than \SI{3}{MeV} will be considered in future work because pair creation should be treated additionally.
Then, we tested the performance for sets of monoenergetic events whose energies are \SI{0.7}{MeV}, \SI{1.0}{MeV}, \SI{1.4}{MeV}, \SI{2.0}{MeV}, and \SI{2.8}{MeV}.
The numbers of training and validation data are summarized in Table \ref{tab:data_num}.
\begin{center}
    \begin{threeparttable}[h]
    \caption{The number of events for each hit in the training/validation data sets}
    \label{tab:data_num}
        \begin{tabular}{ccrrrrr}\hline 
             Hit & Train & 
            \multicolumn{5}{c}{Validation\ [MeV]}\\
             & &
             \multicolumn{1}{c}{0.7} &
             \multicolumn{1}{c}{1.0} &
             \multicolumn{1}{c}{1.4} &
             \multicolumn{1}{c}{2.0} &
             \multicolumn{1}{c}{2.8} \\ \hline\hline
             3 & 47M\tnote{a} & 1.4M           & 1.3M  & 1.2M  & 1.0M & 960K \\
             4 & 40M          & 1.4M           & 1.2M  & 1.1M  & 940K  & 840K \\
             5 & 30M          & 950K\tnote{b}  & 930K  & 870K  & 800K  & 750K \\
             6 & 68M          & 430K           & 490K  & 510K  & 510K  & 530K \\
             7 & 30M          & 520K           & 730K  & 860K  & 990K  & 1.2M \\
             8 & 12M          & 110K           & 200K  & 290K  & 400K  & 570K \\ \hline
        \end{tabular}
        \begin{tablenotes}
            \item[a] $10^6$
            \item[b] $10^3$
        \end{tablenotes}
    \end{threeparttable}
\end{center}

\begin{figure}[htbp]
    \centering
    \includegraphics[width=7cm]{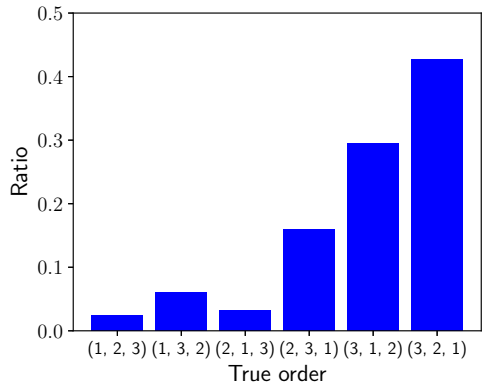}
    \caption{Distribution of the hit order labels for the three-hit events. See the text for meaning of the labels.}
    \label{fig:label_distribution}
\end{figure}

Figure \ref{fig:label_distribution} shows the distribution of hit order labels of three-hit events. The hit index is assigned in the ascending order of the energy deposit as $E_1\leq E_2 \leq E_3$. An order label of $(2, 3, 1)$ means that the first, second, and third hits in the time sequence are labeled by 2, 3, and 1, respectively.
It is known that if the event number of each label in training data are not almost the same, performance of a neural network model is sometimes degraded\cite{Sun2009}.
This distribution bias increases as $n$ gets larger.
We correct the imbalance of the hit order distribution by the following two steps. It should be stressed that this process is applied only to the training data, and the data for performance verification is never modified.

The first step is to eliminate events whose hit order labels rarely appear in original training data. Let $N_i$ be the number of events whose hit order label is $i$-th. Events whose hit order labels are $i$-th are eliminated from the data set if $N_i$ satisfies:
\begin{equation}
    \label{eq:downsampling}
    N_i < \frac{1}{10} \frac{N_{\mathrm{Total}}}{P(n,\ 3)}\ ,
\end{equation}
where $N_{\mathrm{Total}}$ is the number of all the events in the training data. 
This process excludes a part of the data space from being trained, which corresponds to uncommon cases.
This step allows our model to focus on classifying the essential labels for practical imaging and spectroscopy.
Note that this process is independent of the neural network model and is effective only when the distribution of data labels is strongly biased.
We define $M(\leq P(n,3))$ as the number of hit order labels that do not satisfy Equation (\ref{eq:downsampling}).
For example, $N_{\mathrm{min}}$ is less than five hundred in seven-hit events and zero in eight-hit ones in training data we prepare in this work though forty billion photons up to 3.0 MeV are simulated.
Thus, this step prevents the following downsampling from causing a severe shortage of training data.
Table \ref{tab:trim_data} shows the number of discarded hit order labels due to the first step in this simulation data. The fractions of the excluded events are at most 2\% from three to eight-hit ones.
\begin{table}[htbp]
    \centering
    \caption{Number of hit order labels before and after the first step}
    \label{tab:trim_data}
    \begin{tabular}{rcr}\\ \hline
        Hit & Label number & Fraction of excluded events \\ \hline
        3 & 6 to 6 & 0\% \\
        4 & 24 to 24 & 0\% \\
        5 & 60 to 40 & 2\% \\
        6 & 120 to 70 & 2\% \\
        7 & 210 to 117 & 2\% \\
        8 & 336 to 170 & 2\% \\ \hline
    \end{tabular}
\end{table}

The second step performs downsampling. Downsampling is one of the powerful solutions for multi-class classification of imbalanced data sets\cite{Batista2004}. Suppose aggregating the simulated events in terms of the hit order and counting event number of each label as $\{N_i\}=(N_1,N_2,\ldots,N_{M})$. 
$N_{\mathrm{min}}$ is defined as $\min\, \{N_i\}$. 
Downsampling means utilizing randomly chosen $N_{\mathrm{min}}$ events for each hit order label in training.
Thanks to this step, the event number of each label in a minibatch of one training step is the same, which prompts efficient training.

\section{Results}
\label{sec:result}
After training the neural network model, we reconstructed the monoenergetic evaluation data. After evaluating angular and energy resolutions, we study the effects of the number of hidden layers.
In addition, we compare the performance of the neural network with those of two other algorithms: a conventional, classical model and a physics-based probabilistic one.

\subsection{Distribution of ARM and reconstructed energy}
\label{sec:result_1}
\begin{figure*}[ht]
    \centering
    \begin{tabular}{c}
        \begin{minipage}{0.33\hsize}
            \centering
            \includegraphics[width=5.15cm]{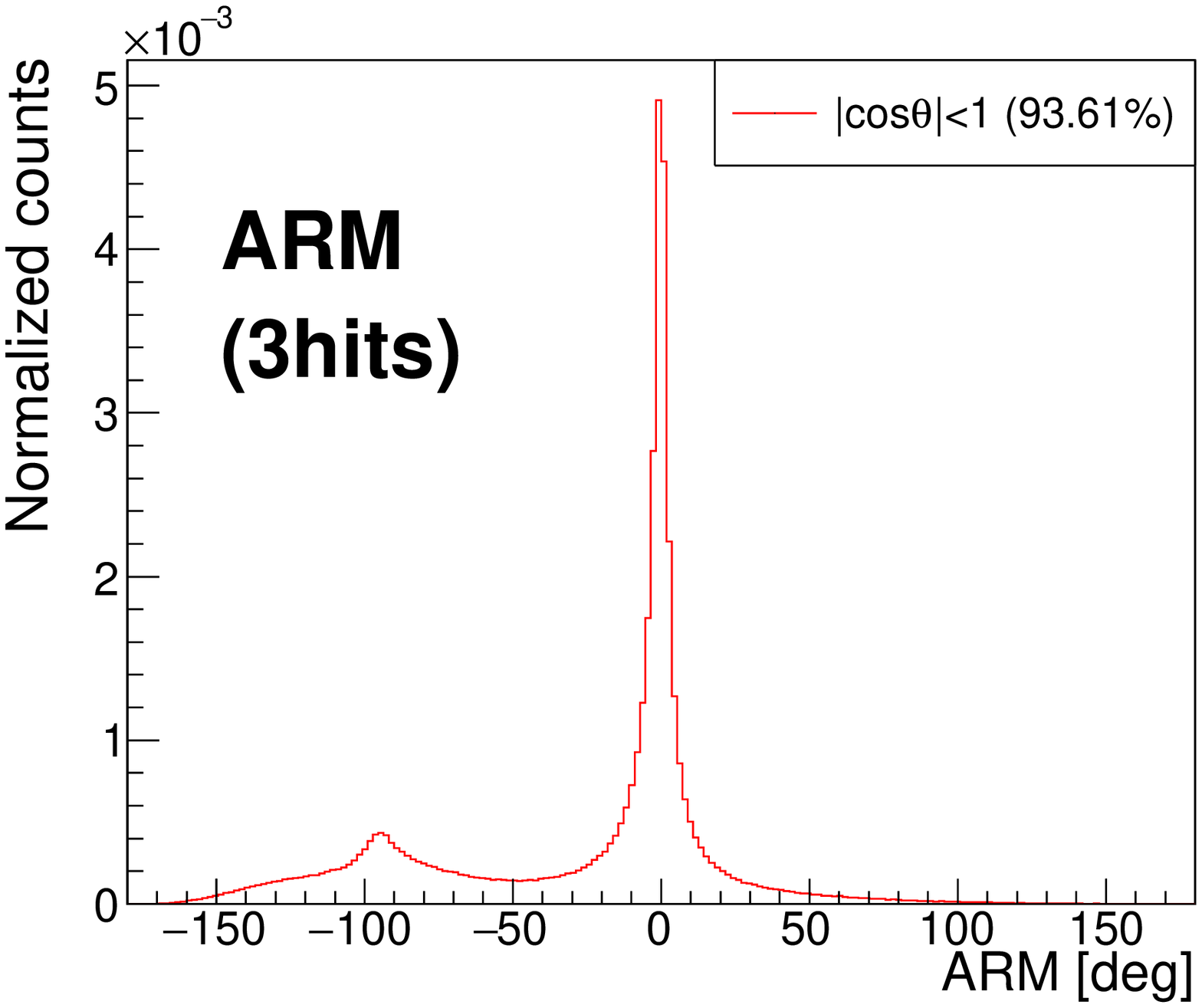}
        \end{minipage}
        \begin{minipage}{0.33\hsize}
            \centering
            \includegraphics[width=5.15cm]{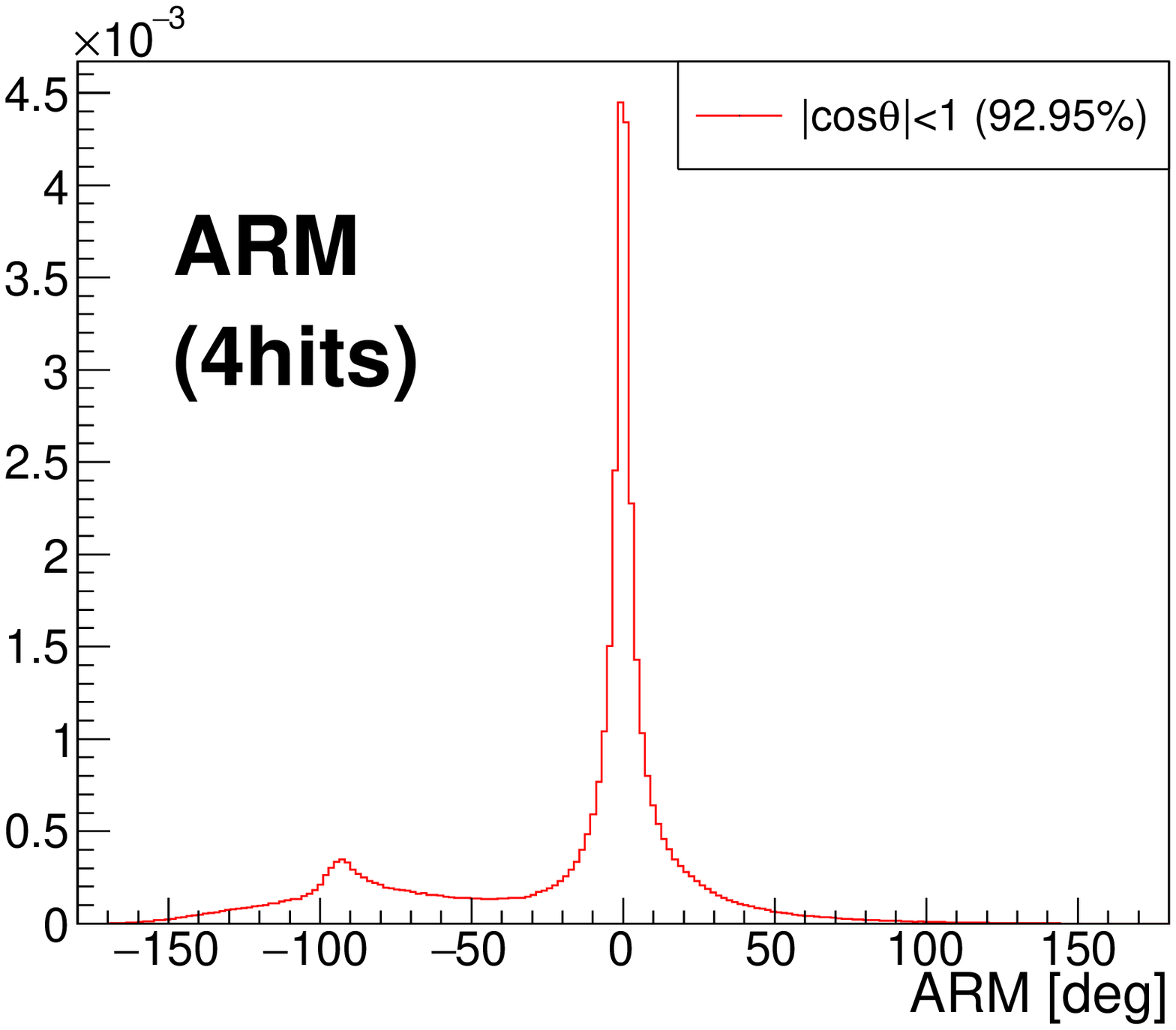}
        \end{minipage}
        \begin{minipage}{0.33\hsize}
            \centering
            \includegraphics[width=5.15cm]{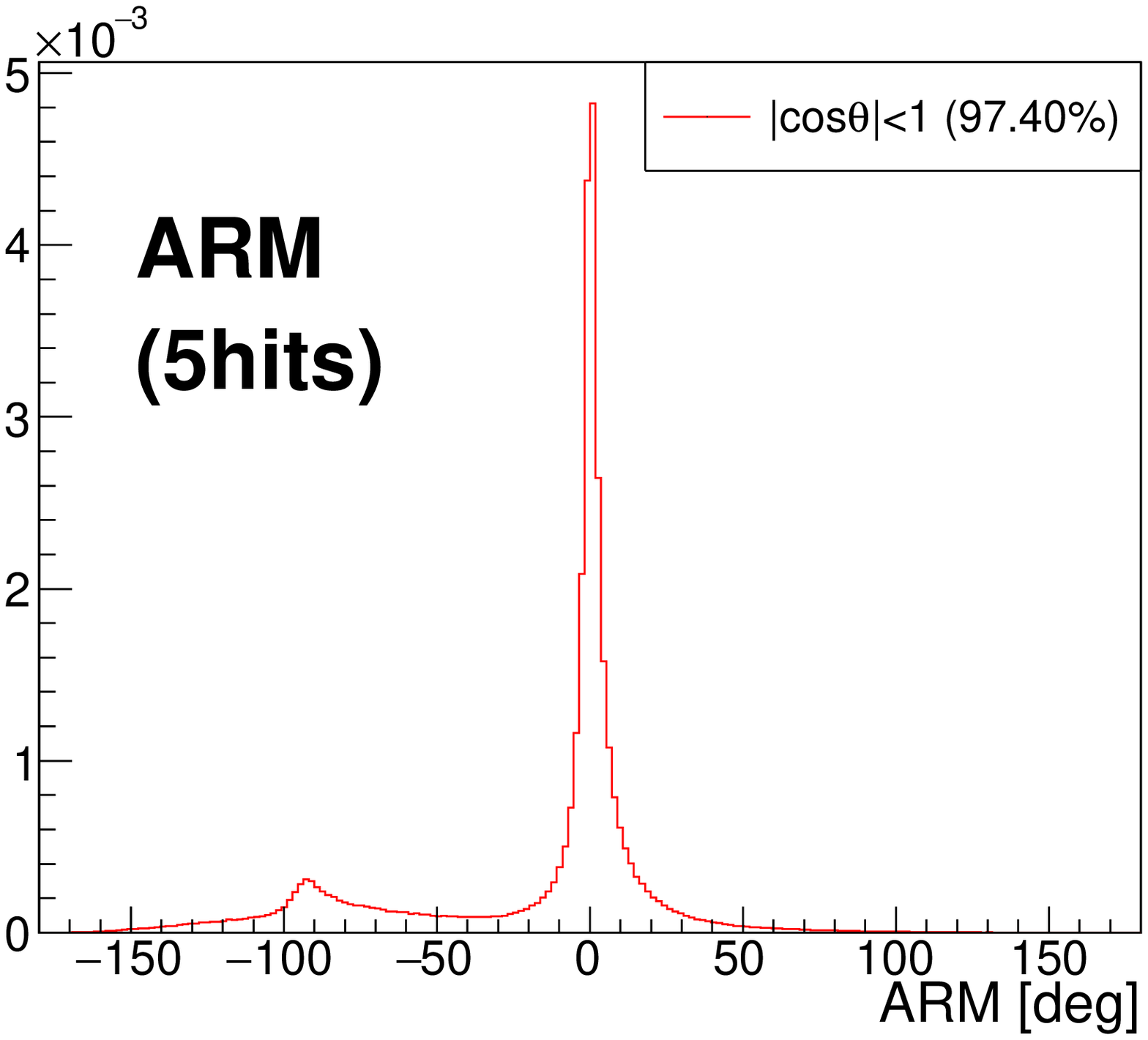}
        \end{minipage}\\
        \begin{minipage}{0.33\hsize}
            \centering
            \includegraphics[width=5.15cm]{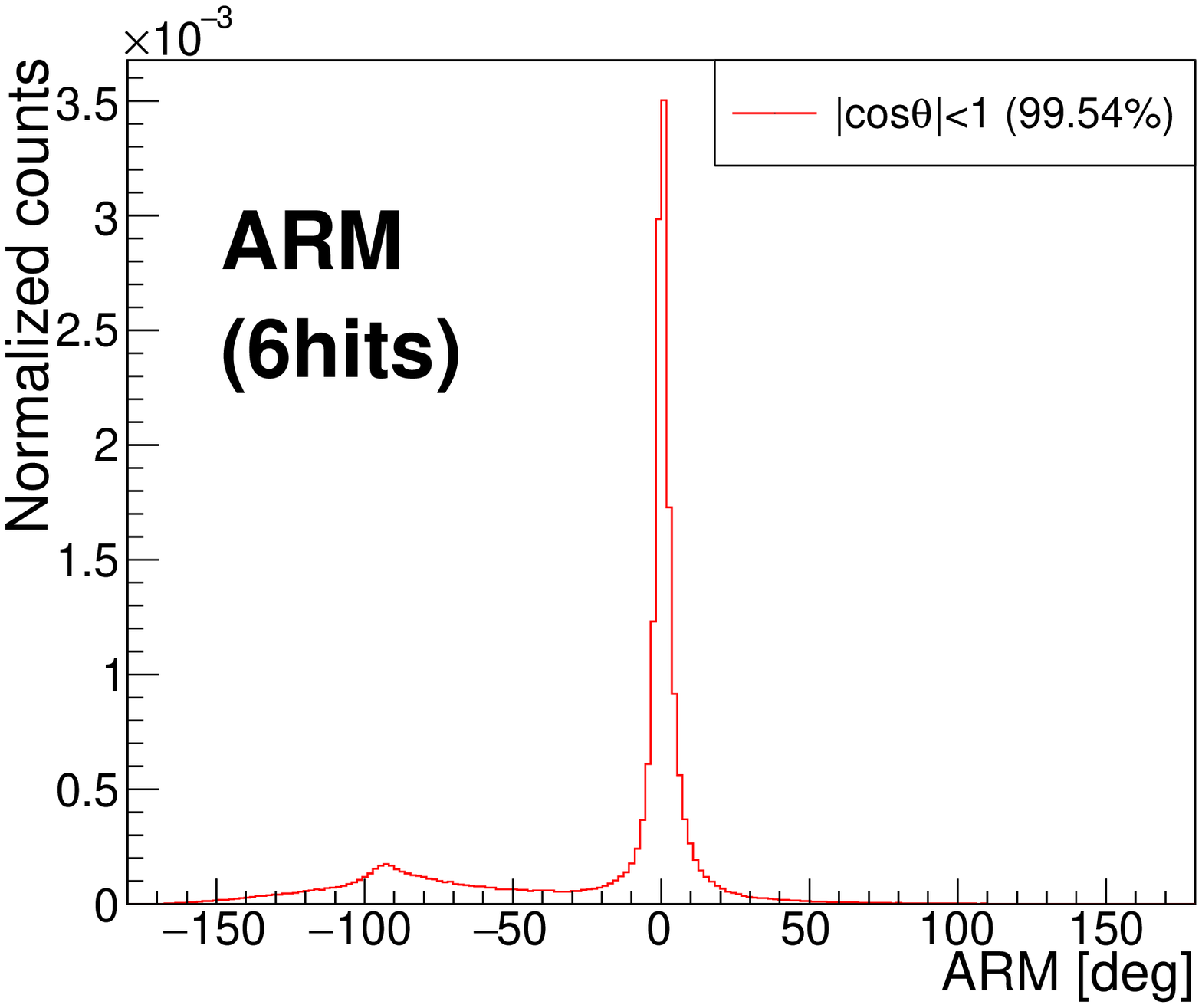}
        \end{minipage}
        \begin{minipage}{0.33\hsize}
            \centering
            \includegraphics[width=5.15cm]{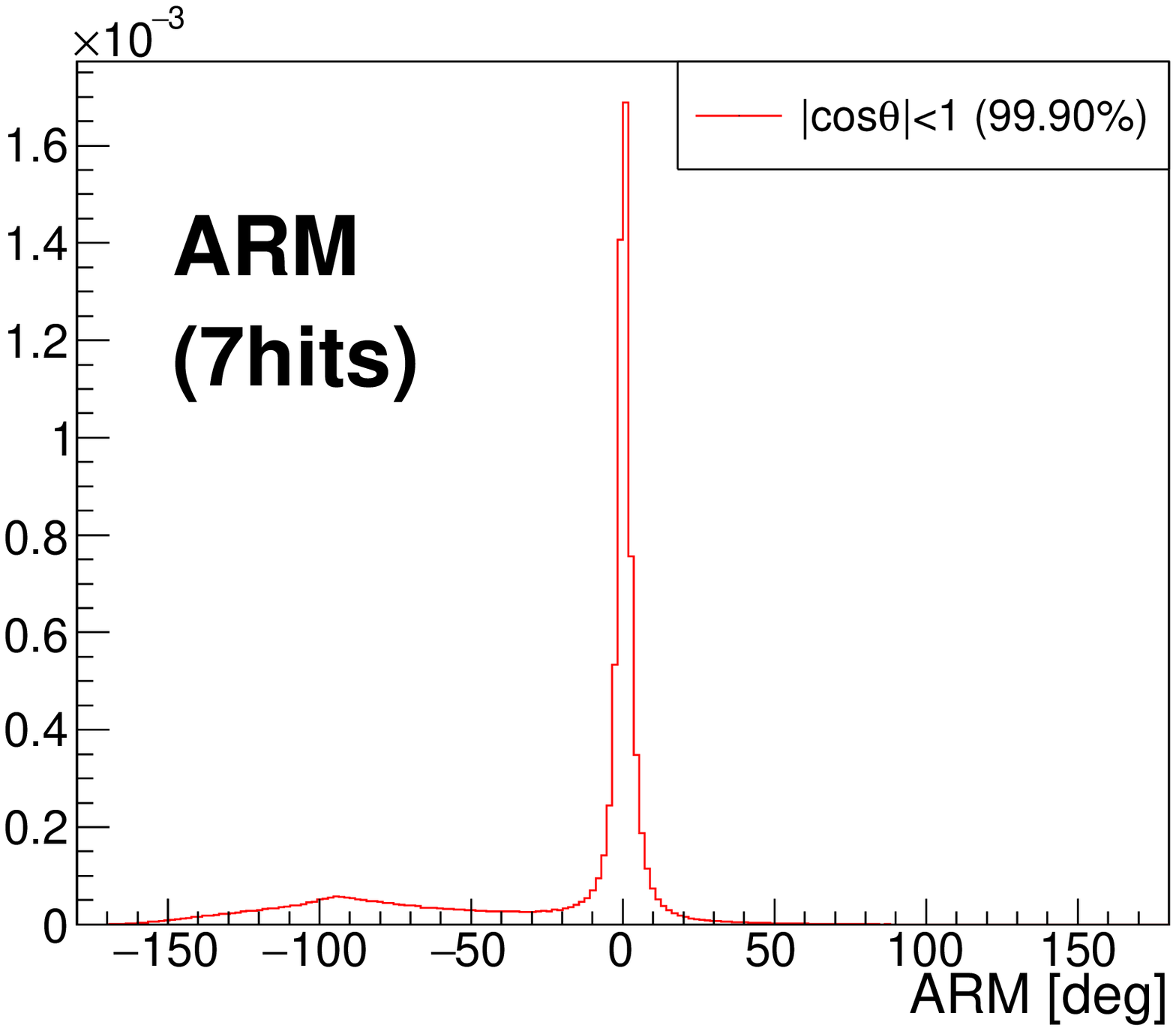}
        \end{minipage}
        \begin{minipage}{0.33\hsize}
            \centering
            \includegraphics[width=5.15cm]{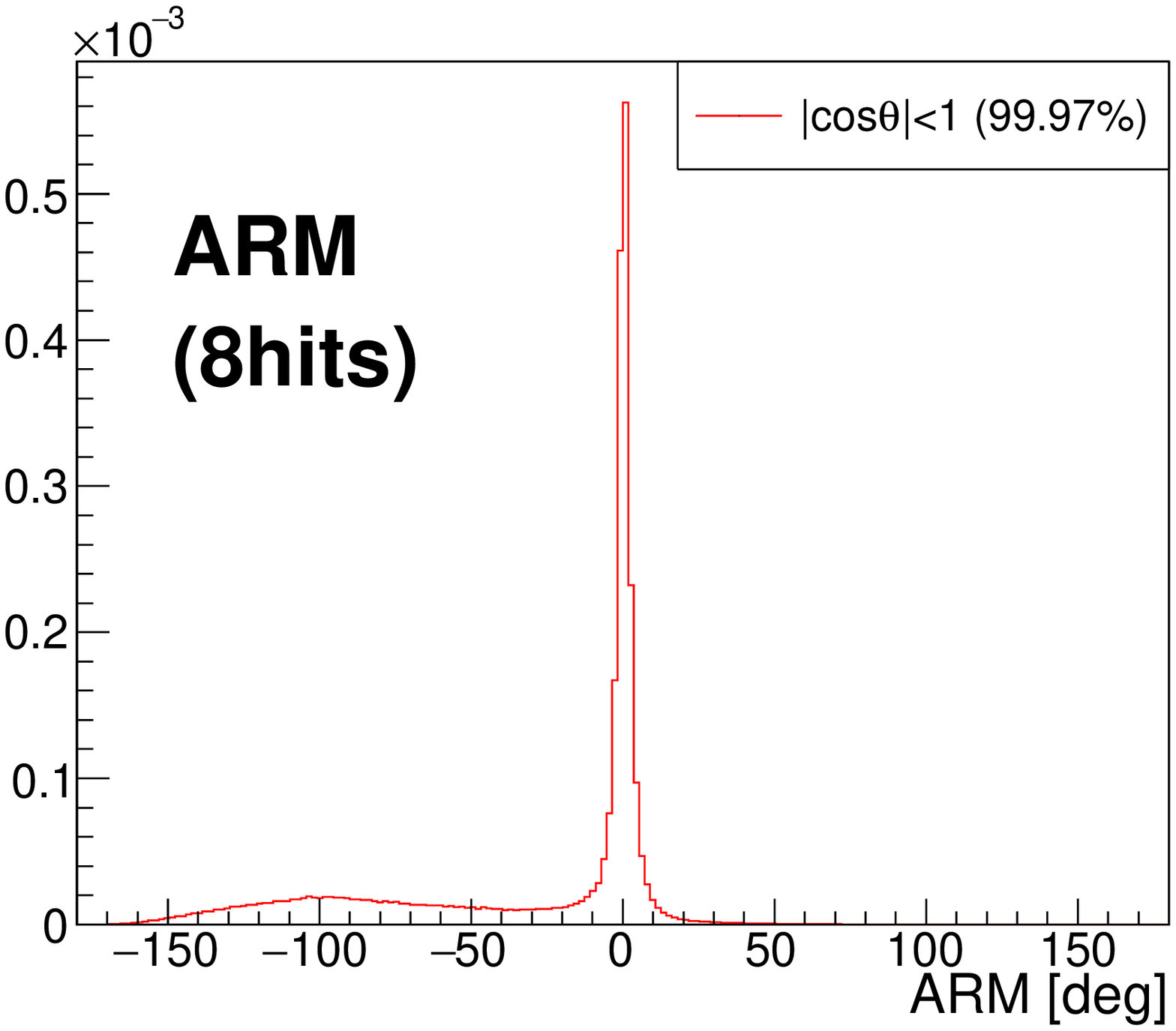}
        \end{minipage}
    \end{tabular}
    \caption{ARM distributions of reconstructed \SI{1.4}{MeV} photon events. Only good events whose first scattering angle $\theta$ satisfies $-1<\cos{\theta}<1$ are extracted. The fractions of good events to all the events are noted on each legend.}
    \label{fig:arm_results}
\end{figure*}

\begin{figure*}[ht]
    \begin{tabular}{c}
        \begin{minipage}{0.33\hsize}
            \centering
            \includegraphics[width=5.15cm]{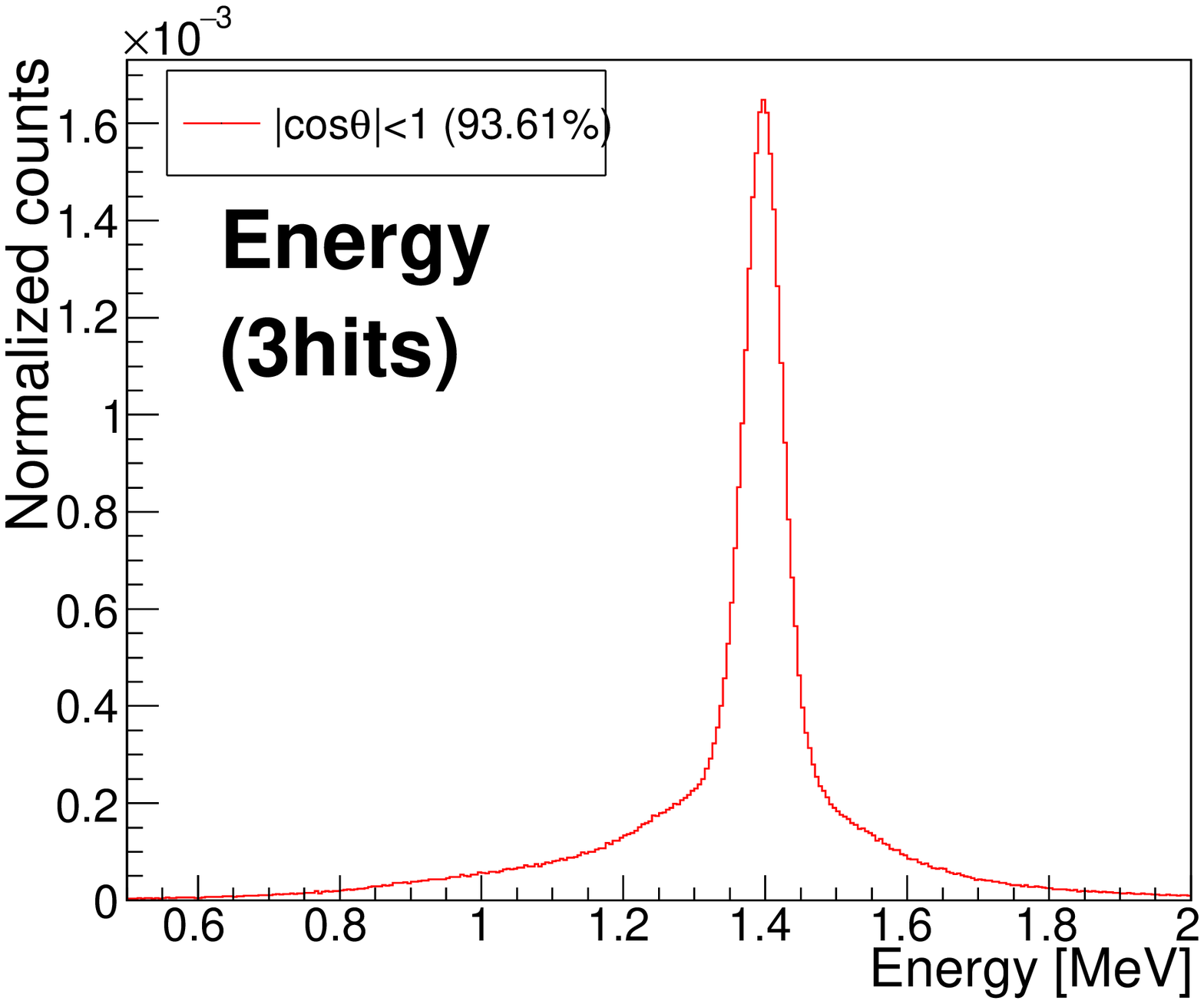}
        \end{minipage}
        \begin{minipage}{0.33\hsize}
            \centering
            \includegraphics[width=5.15cm]{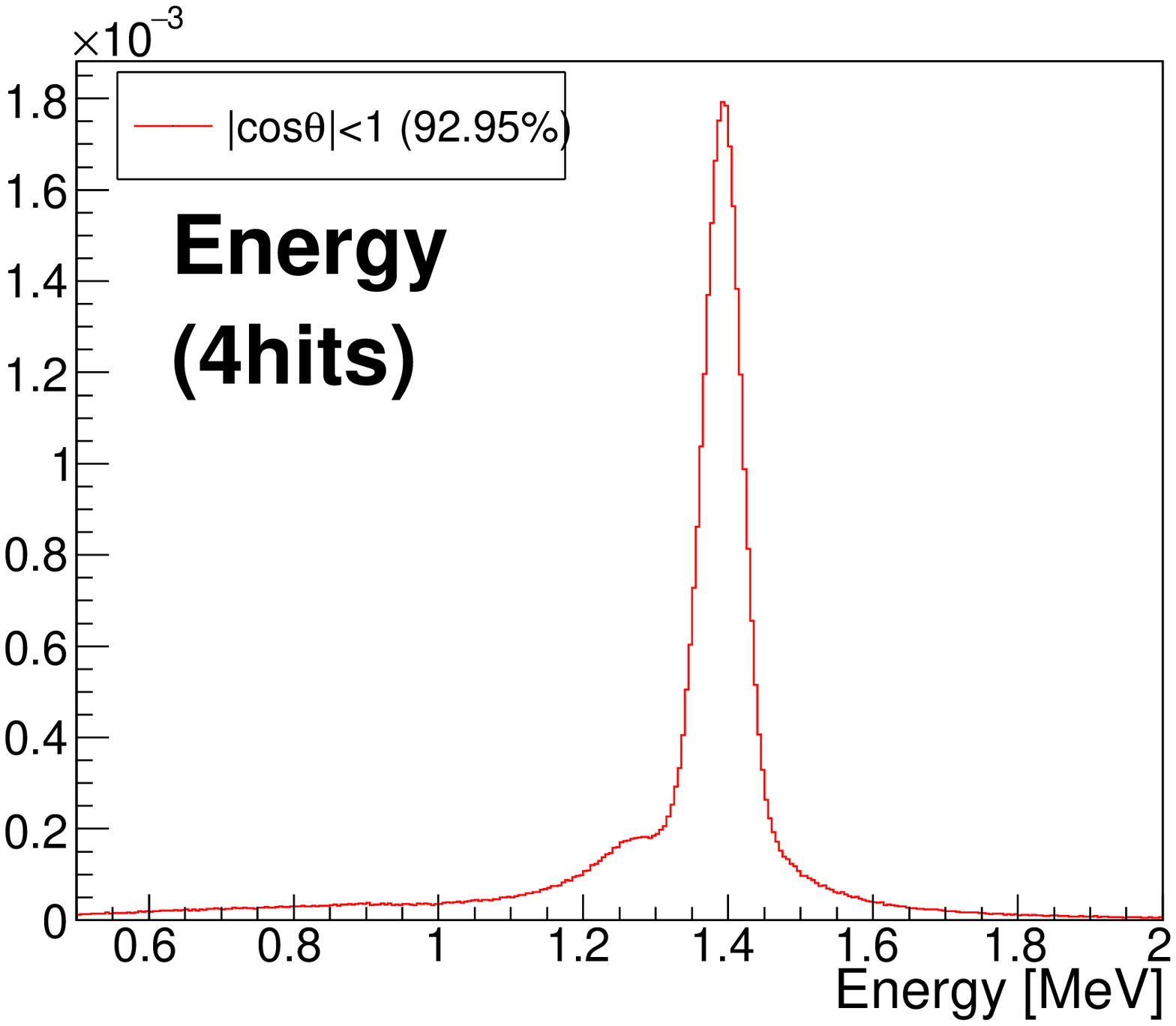}
        \end{minipage}
        \begin{minipage}{0.33\hsize}
            \centering
            \includegraphics[width=5.15cm]{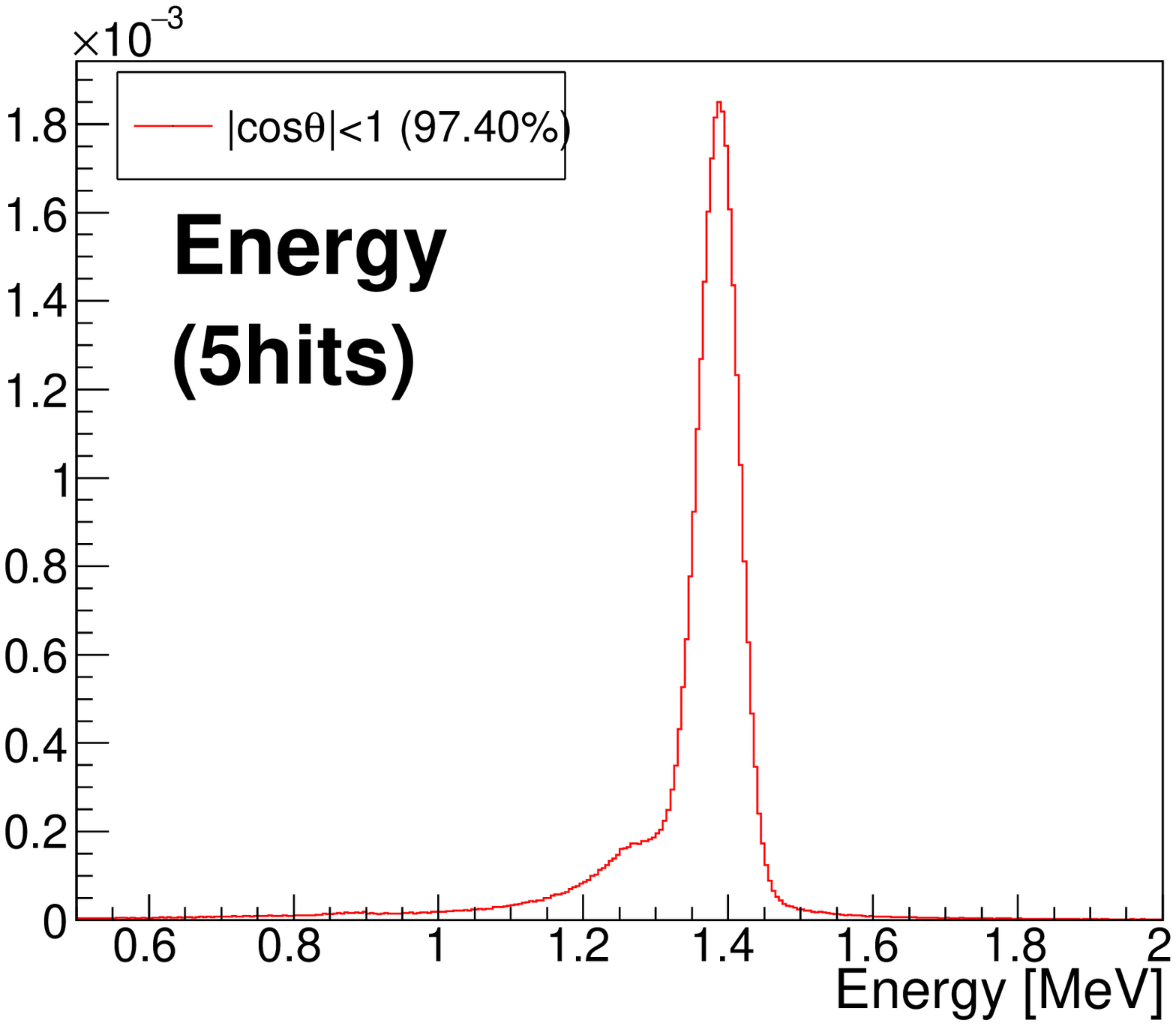}
        \end{minipage}\\
        \begin{minipage}{0.33\hsize}
            \centering
            \includegraphics[width=5.15cm]{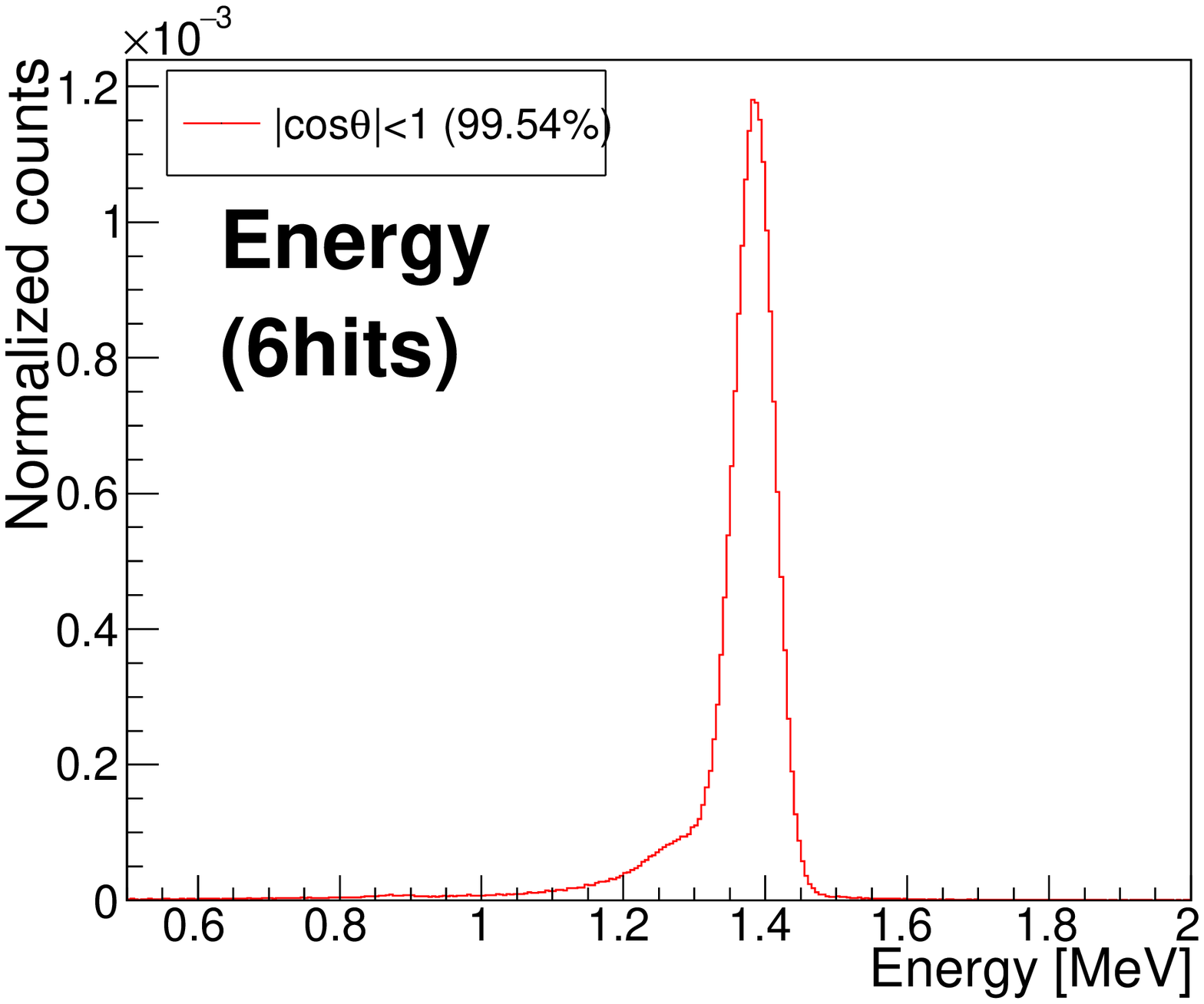}
        \end{minipage}
        \begin{minipage}{0.33\hsize}
            \centering
            \includegraphics[width=5.15cm]{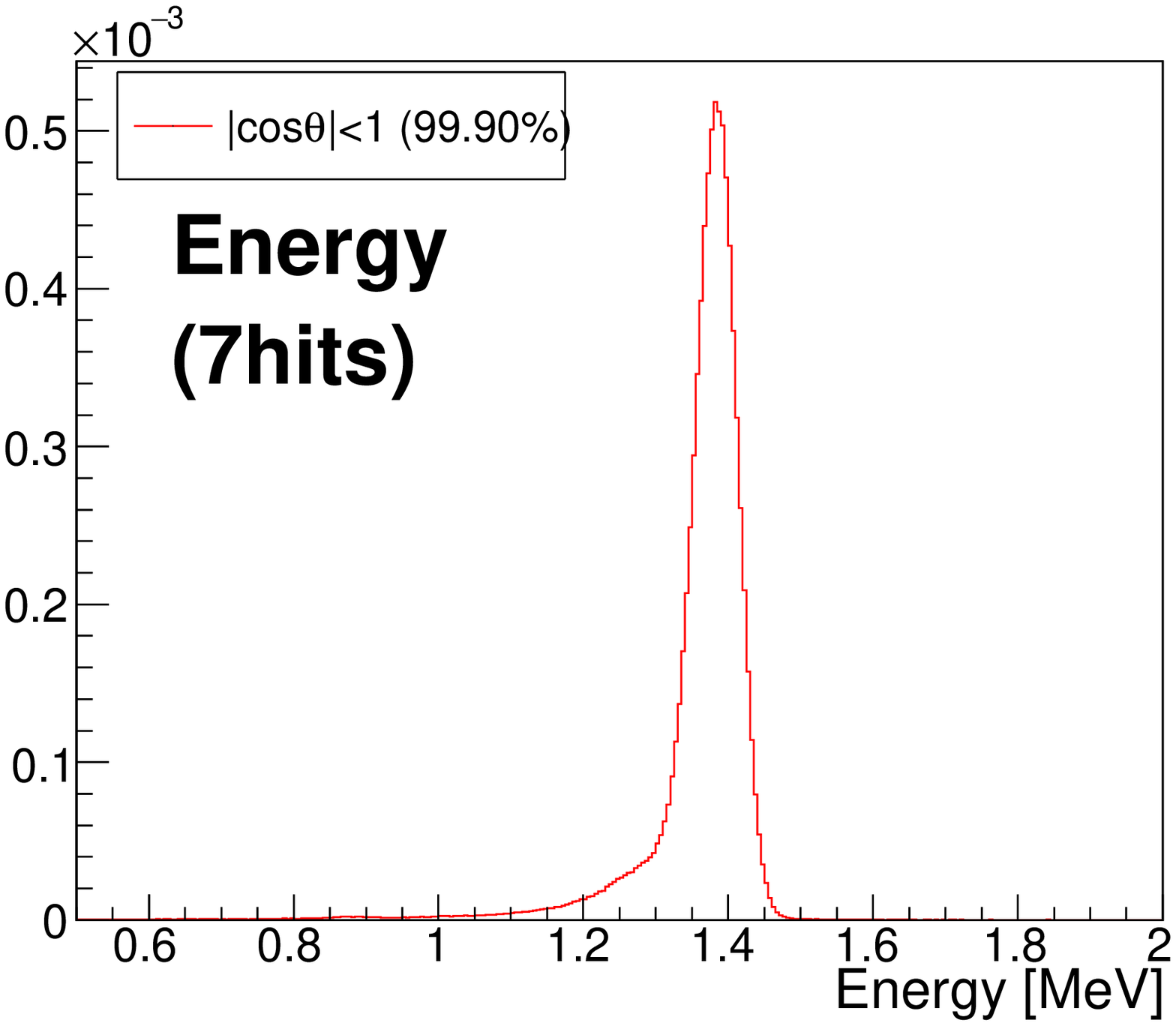}
        \end{minipage}
        \begin{minipage}{0.33\hsize}
            \centering
            \includegraphics[width=5.15cm]{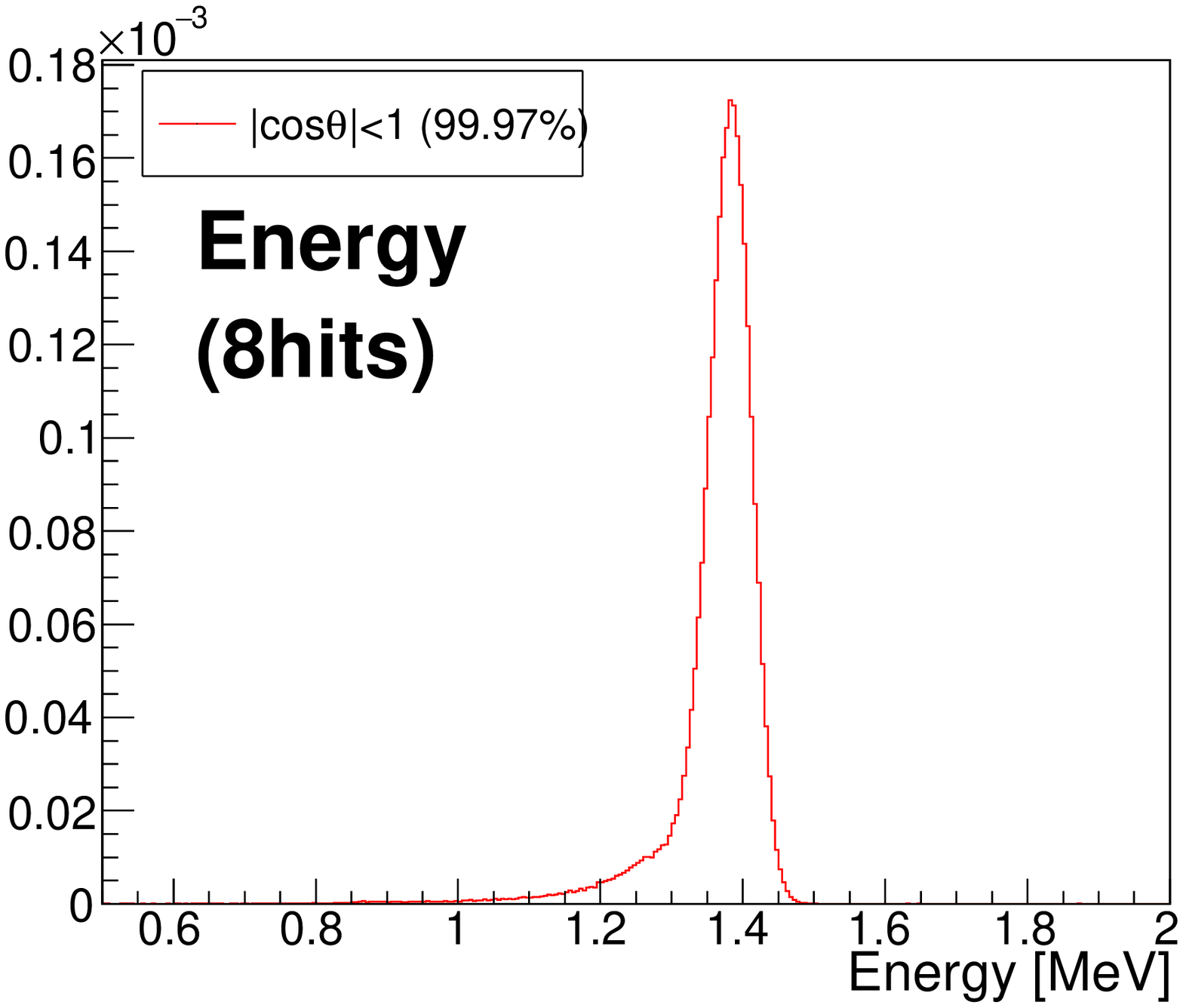}
        \end{minipage}
    \end{tabular}
    \caption{Energy distributions of reconstructed \SI{1.4}{MeV} photon events. Many reconstructed events are distributed around \SI{1.4}{MeV}, which shows excellent performance of event reconstruction. Meanings of fractions in the legends are the same as Figure \ref{fig:arm_results}.}
    \label{fig:energy_results}
\end{figure*}

Distributions of the ARM and the reconstructed energy directly reflect performance of a Compton telescope.
Figure \ref{fig:arm_results} shows ARM distributions for the reconstructed \SI{1.4}{MeV} photon events.
The fractions of events that satisfy $-1 < \cos\theta_{\mathrm{K}} < 1$ are also shown in this figure; note that $\cos\theta_{\mathrm{K}}$ calculated by Equation (\ref{eq:compton}) can be out of $(-1,1)$ due to an incorrect reconstruction, finite position and energy resolutions.
These fractions are higher than \SI{92}{\%} for the three to eight-hit events.
The sharp peaks around zero manifest a good performance of the event reconstruction.
Figure \ref{fig:energy_results} shows reconstructed energy spectra for the same data set.
Again, the sharp peaks around \SI{1.4}{MeV} show that the initial energy is correctly estimated.

\begin{figure*}[ht]
    \begin{tabular}{c}
        \begin{minipage}{0.33\hsize}
            \centering
            \includegraphics[width=5.15cm]{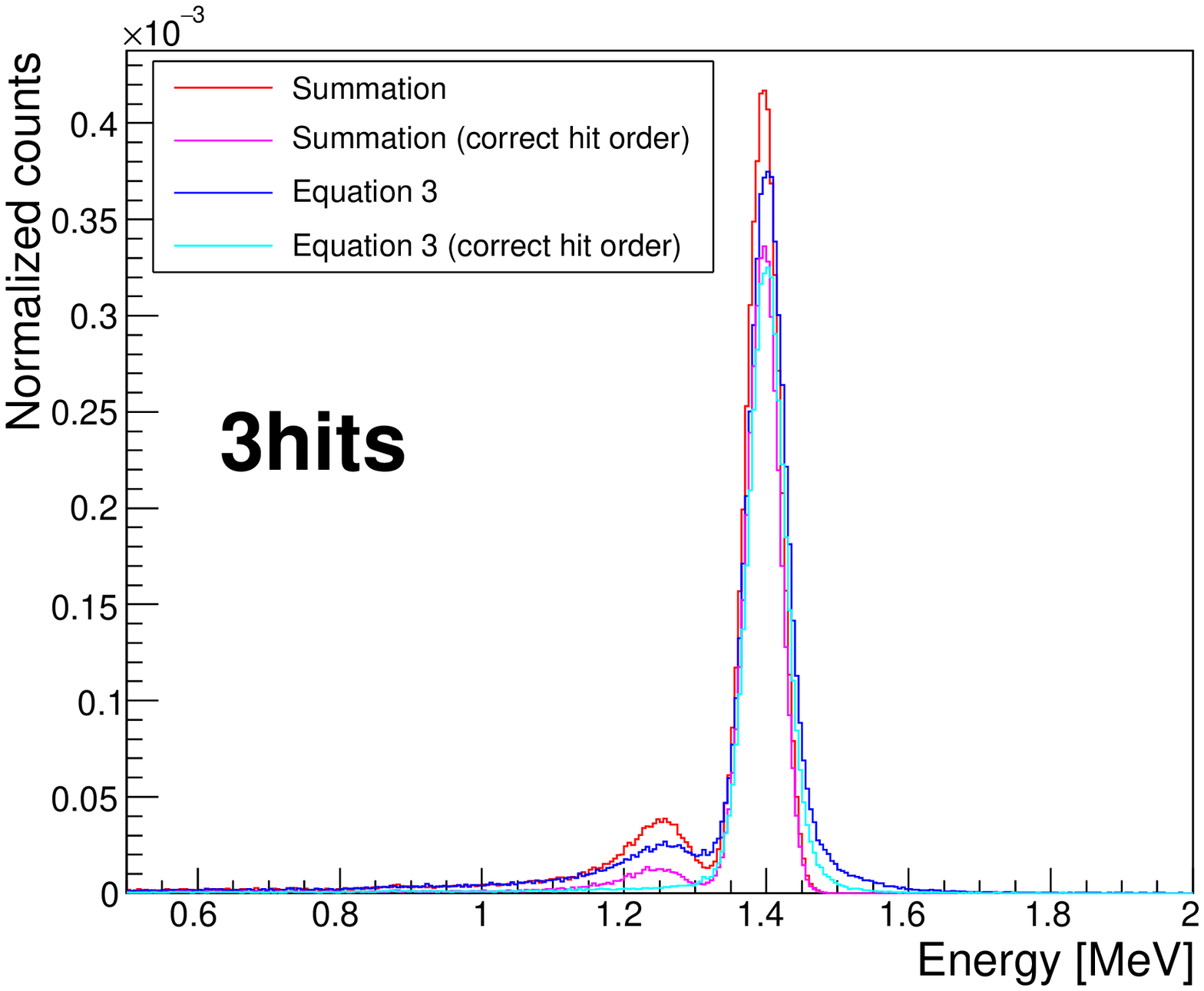}
        \end{minipage}
        \begin{minipage}{0.33\hsize}
            \centering
            \includegraphics[width=5.15cm]{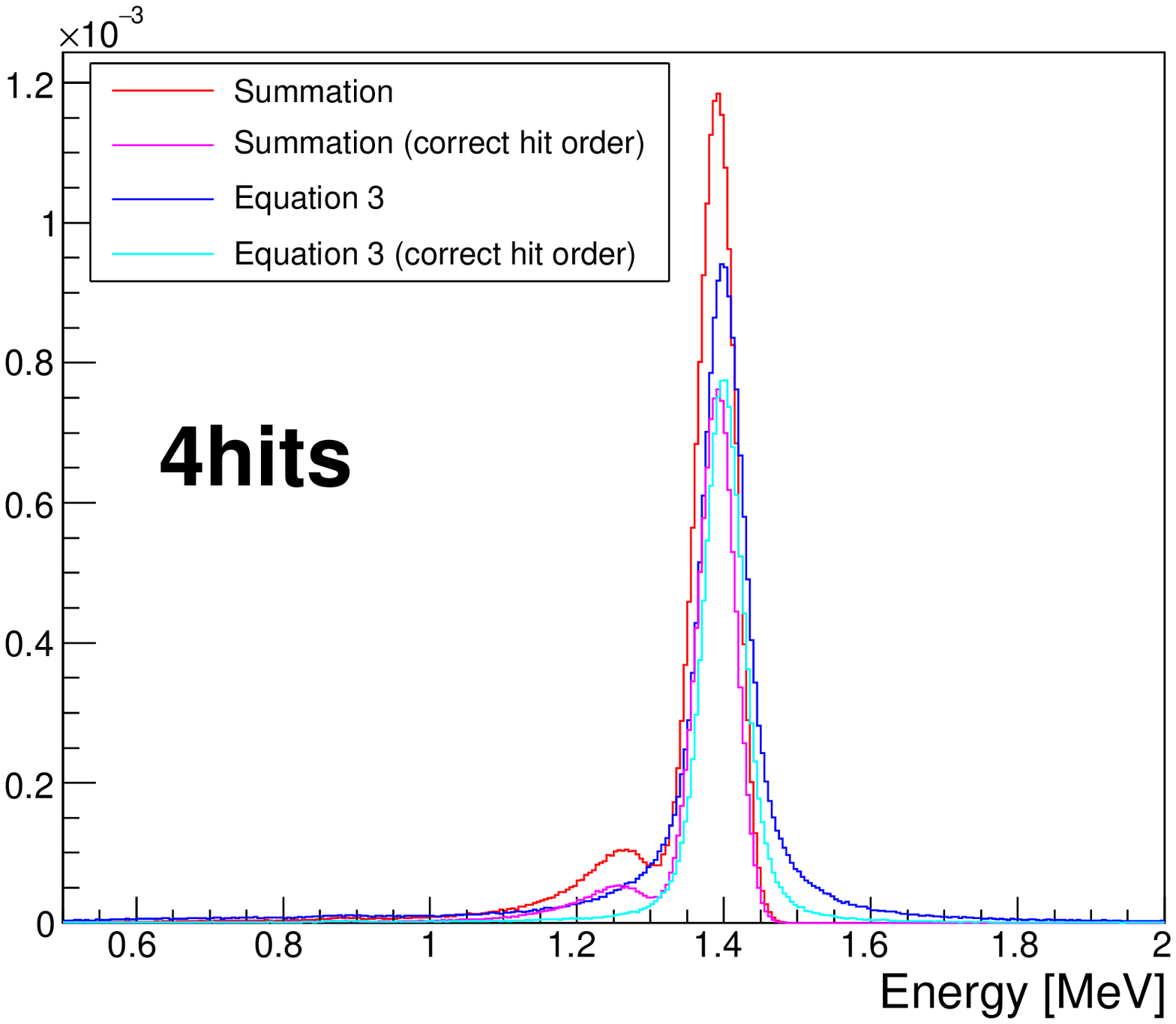}
        \end{minipage}
        \begin{minipage}{0.33\hsize}
            \centering
            \includegraphics[width=5.15cm]{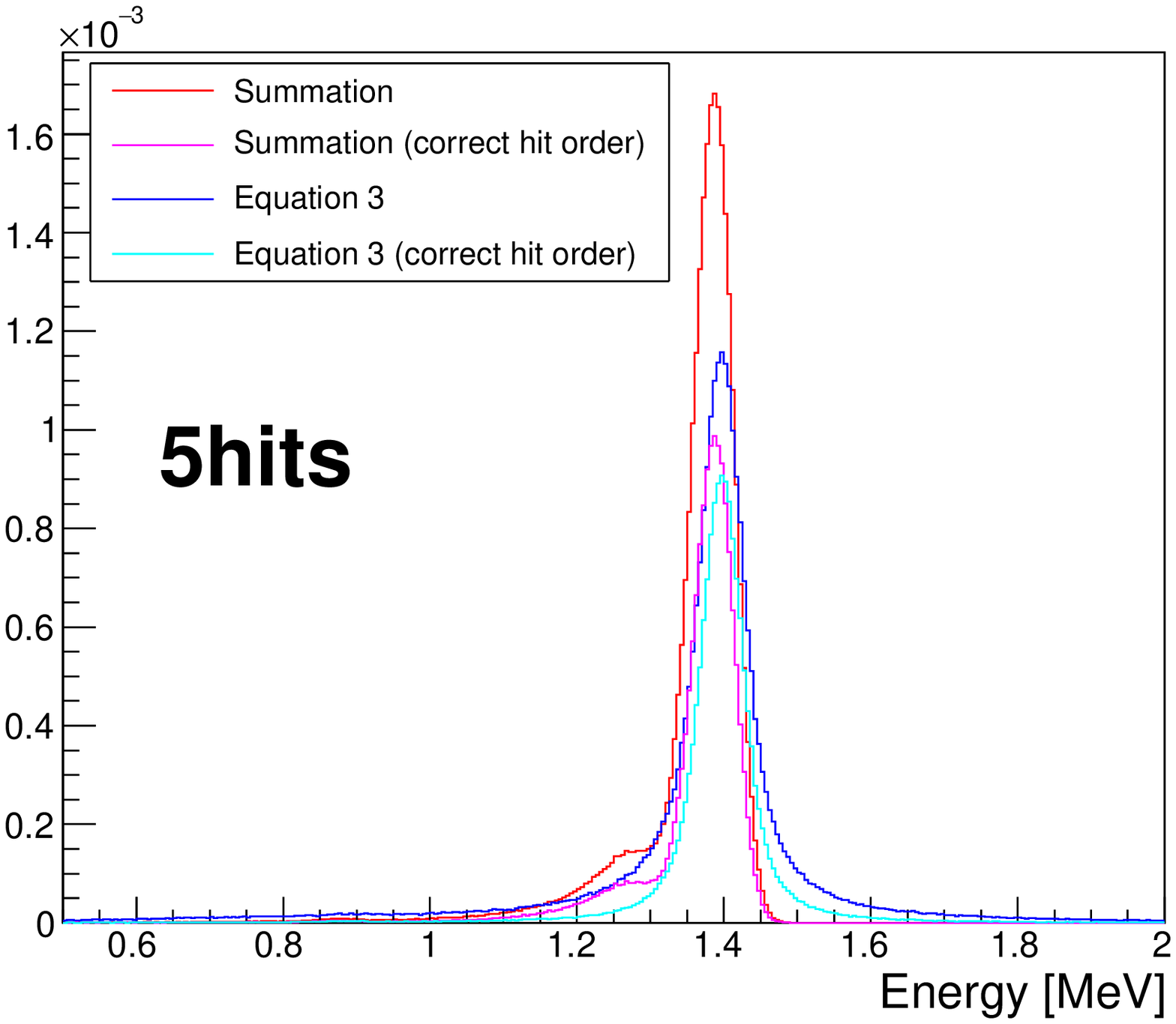}
        \end{minipage}\\
        \begin{minipage}{0.33\hsize}
            \centering
            \includegraphics[width=5.15cm]{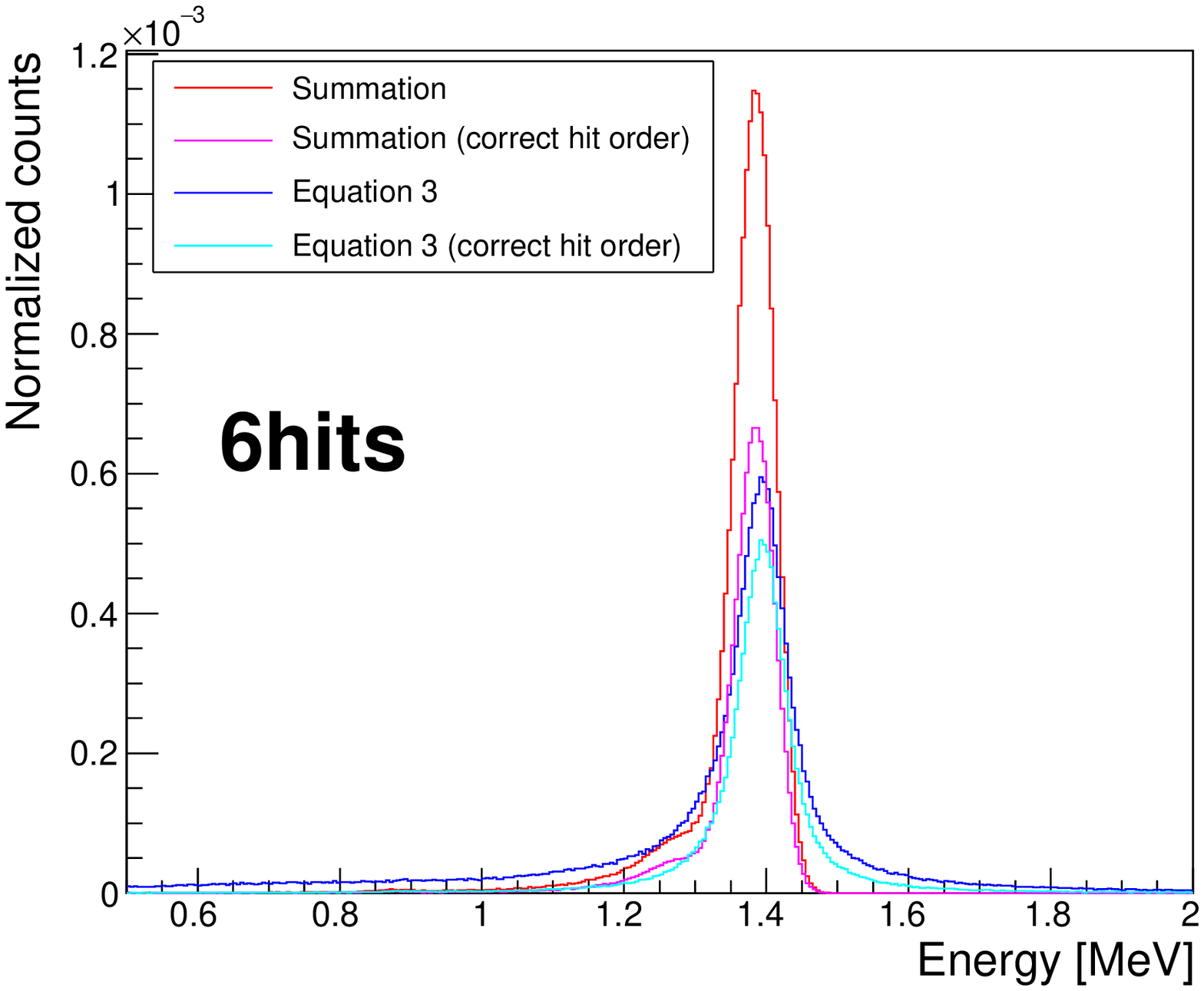}
        \end{minipage}
        \begin{minipage}{0.33\hsize}
            \centering
            \includegraphics[width=5.15cm]{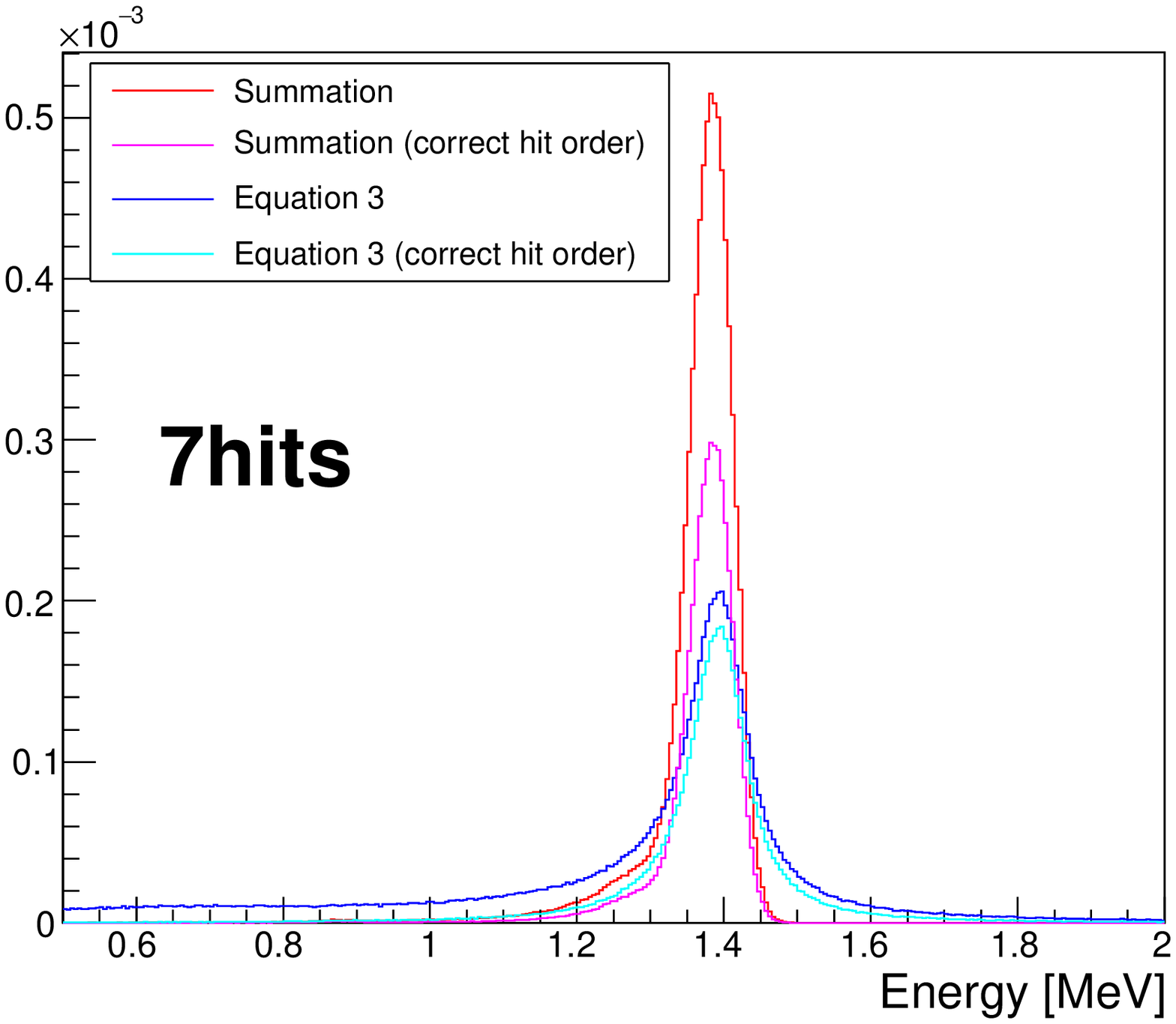}
        \end{minipage}
        \begin{minipage}{0.33\hsize}
            \centering
            \includegraphics[width=5.15cm]{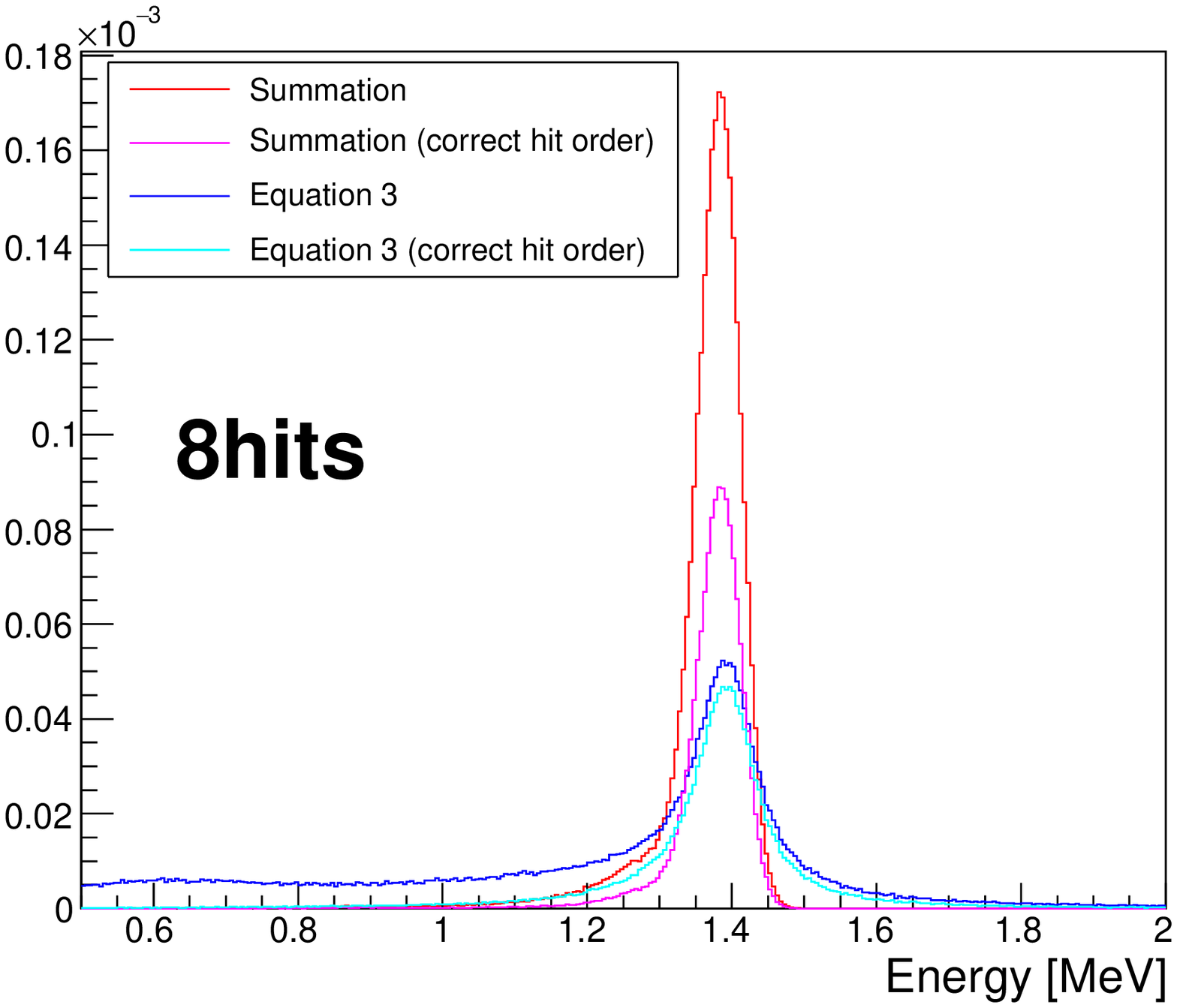}
        \end{minipage}
    \end{tabular}
    \caption{Reconstructed energy of \SI{1.4}{MeV} monoenergetic photons predicted as fully-absorbed events. Red lines denote the results calculated from summation of all the energies, which include events with wrong hit order labels. Magenta ones do the same but limited to events with correct order labels. Blue and light blue lines are those using Equation (\ref{eq:estimated_energy_escape}). In all case, the reconstruction by the summation shows better resolutions even for events whose hit order labels are correct.}
    \label{fig:escape_flag_effect}
\end{figure*}

As described in Section \ref{sec:model_1}, it is appropriate to calculate the initial energy by summing all energy deposits of the hits in the case of a fully-absorbed event.
If the first three-hit order is correctly determined these estimated energies should be equal to those calculated by Equation (\ref{eq:estimated_energy_escape}).
Figure \ref{fig:escape_flag_effect} shows differences in the distribution of the reconstructed energy between the two ways. 
The uncertainty of the estimated initial energy by Equation (\ref{eq:estimated_energy_escape}) (blue line) is larger than that by summing all energy deposits (red line).
The small accumulated structures lower than \SI{1.3}{MeV} for $n=3$ and $n=4$ is attributed to escape events that are incorrectly predicted as fully-deposit ones.
The shape of this accumulation below the main peak has a long low-energy tail and a relatively sharp upper limit corresponding to the maximum allowed total energy deposit by multiple scatterings.

One reason that the summing method shows sharper peaks of the energy distribution is that the energy estimation is not affected by the wrong order estimation.
Another is an accumulation of propagated errors in Equation (\ref{eq:estimated_energy_escape}).
magenta lines in Figure \ref{fig:escape_flag_effect} show events whose hit orders are correctly determined and whose initial energies are calculated by summing deposits, while light blue ones are the same but Equation (\ref{eq:estimated_energy_escape}) is used for energy estimation.
Those peaks of fully-absorbed events get sharper if the energy is calculated by summation.
It is therefore essential to employ the escape flag and different energy estimation methods according to the predicted value of the flag --- a fully-absorbed or an escape event.

\begin{figure}[htbp]
    \centering
    \includegraphics[width=7cm]{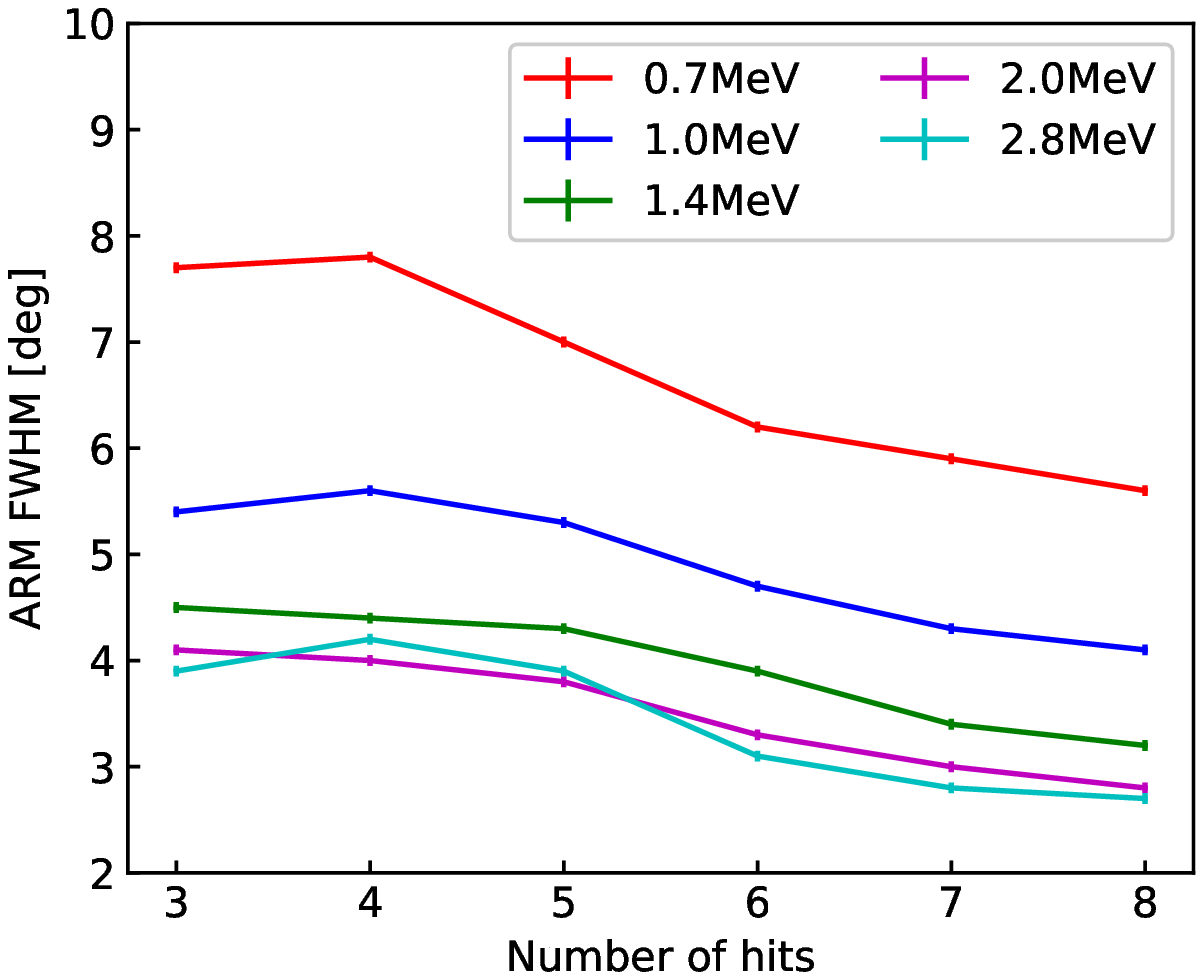}
    \caption{FWHM of ARM distributions in the band from 0.7 to \SI{2.8}{MeV}. Each color corresponds to the different initial energy of events.}
    \label{fig:fwhm_arm_first3hits}
\end{figure}

\begin{figure}[htbp]
    \centering
    \includegraphics[width=7cm]{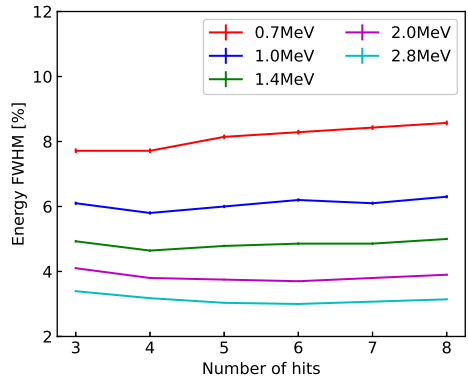}
    \caption{FWHM of reconstructed energy spectra. As the initial energy increases the energy resolution improves like Figure \ref{fig:fwhm_arm_first3hits}.}
    \label{fig:fwhm_energy_first3hits}
\end{figure}

Evaluating the peak widths of the ARM and the reconstructed energy measure the model performance in a quantitative way.
Figure \ref{fig:fwhm_arm_first3hits} shows the angular resolutions in full width at half maximum (FWHM) for several monoenergetic photon events. 
In general, the angular resolution gets better as the hit number and initial energy increase.
The best FWHM of the ARM is \ang{2.7} for eight-hit events in \SI{2.8}{MeV}. Figure \ref{fig:fwhm_energy_first3hits} shows the energy resolution in FWHM.
The energy resolution gets better with the initial energy on the assumption of our detector settings. The degradation of the energy resolution at lower energies is attributed to the measurement error of the energy deposit of each hit by the assumed detector.

\begin{figure*}[ht]
    \begin{tabular}{c}
        \begin{minipage}{0.33\hsize}
            \centering
            \includegraphics[width=5.15cm]{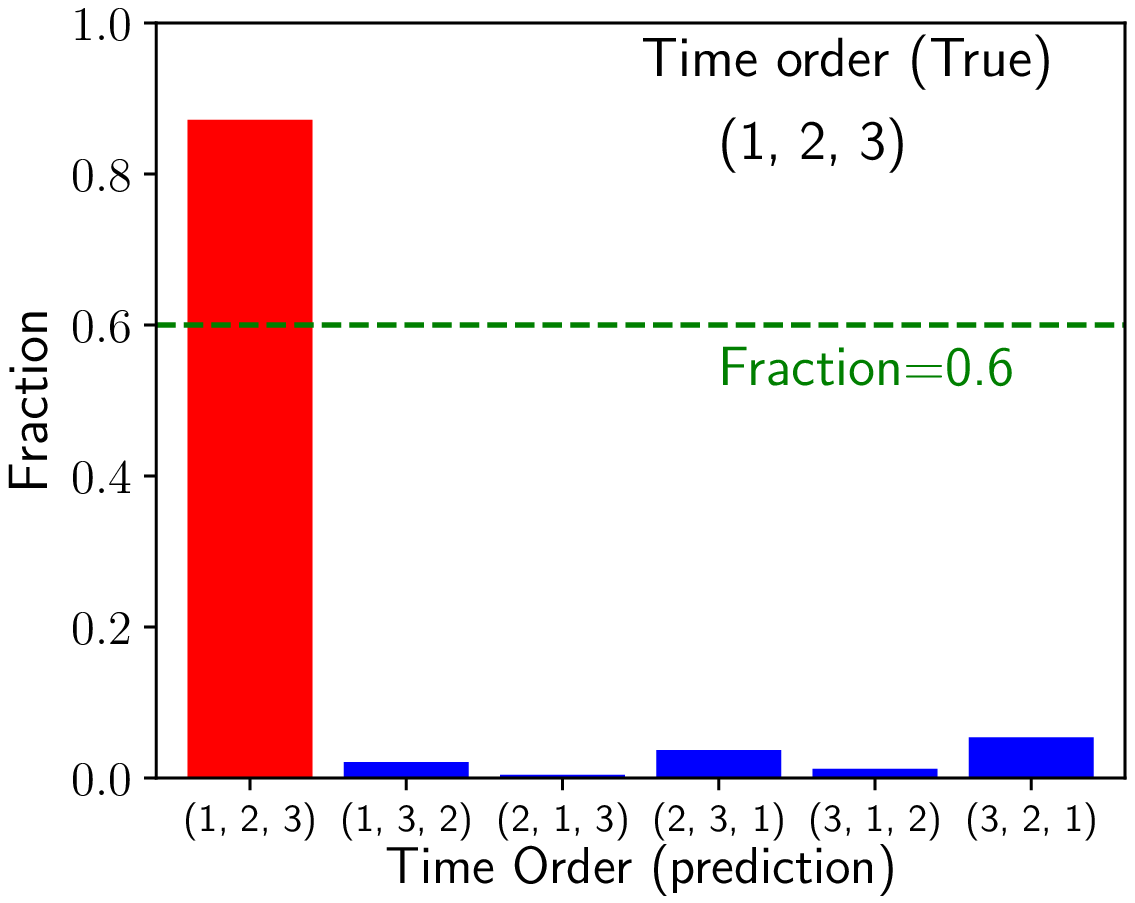}
        \end{minipage}
        \begin{minipage}{0.33\hsize}
            \centering
            \includegraphics[width=5.15cm]{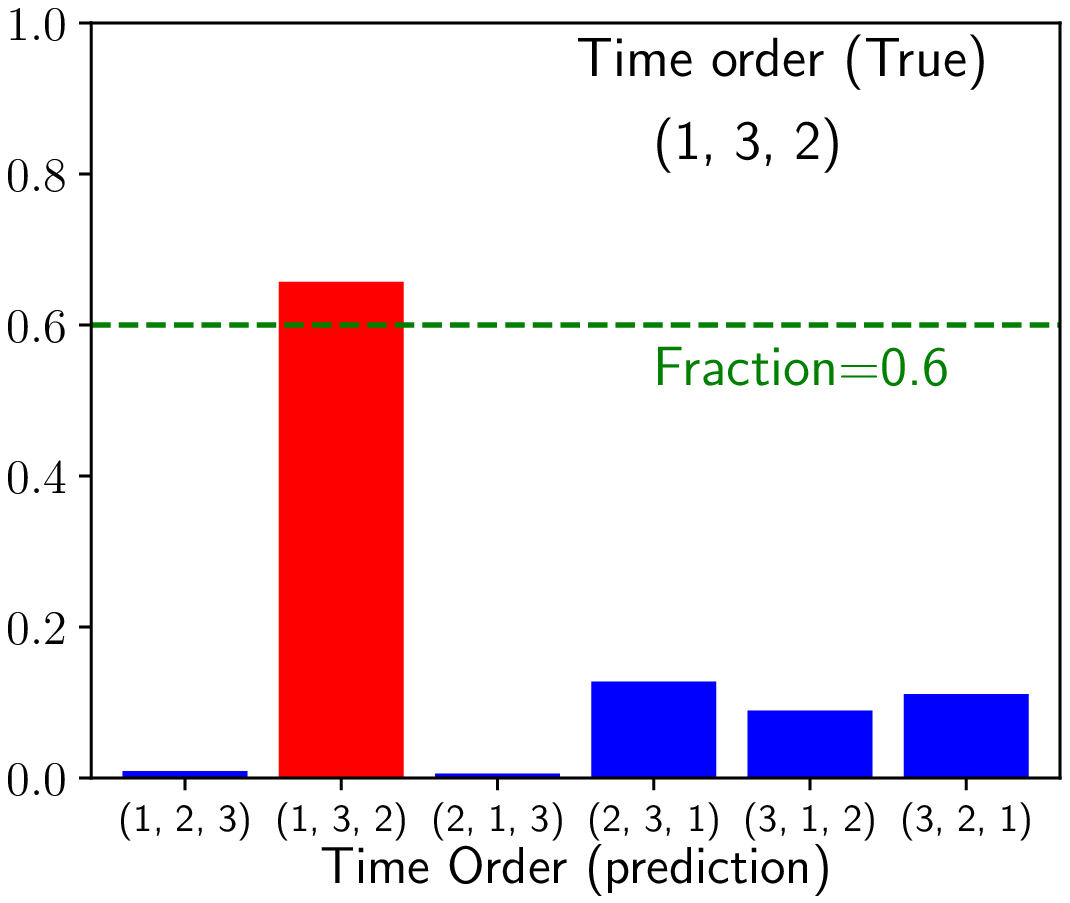}
        \end{minipage}
        \begin{minipage}{0.33\hsize}
            \centering
            \includegraphics[width=5.15cm]{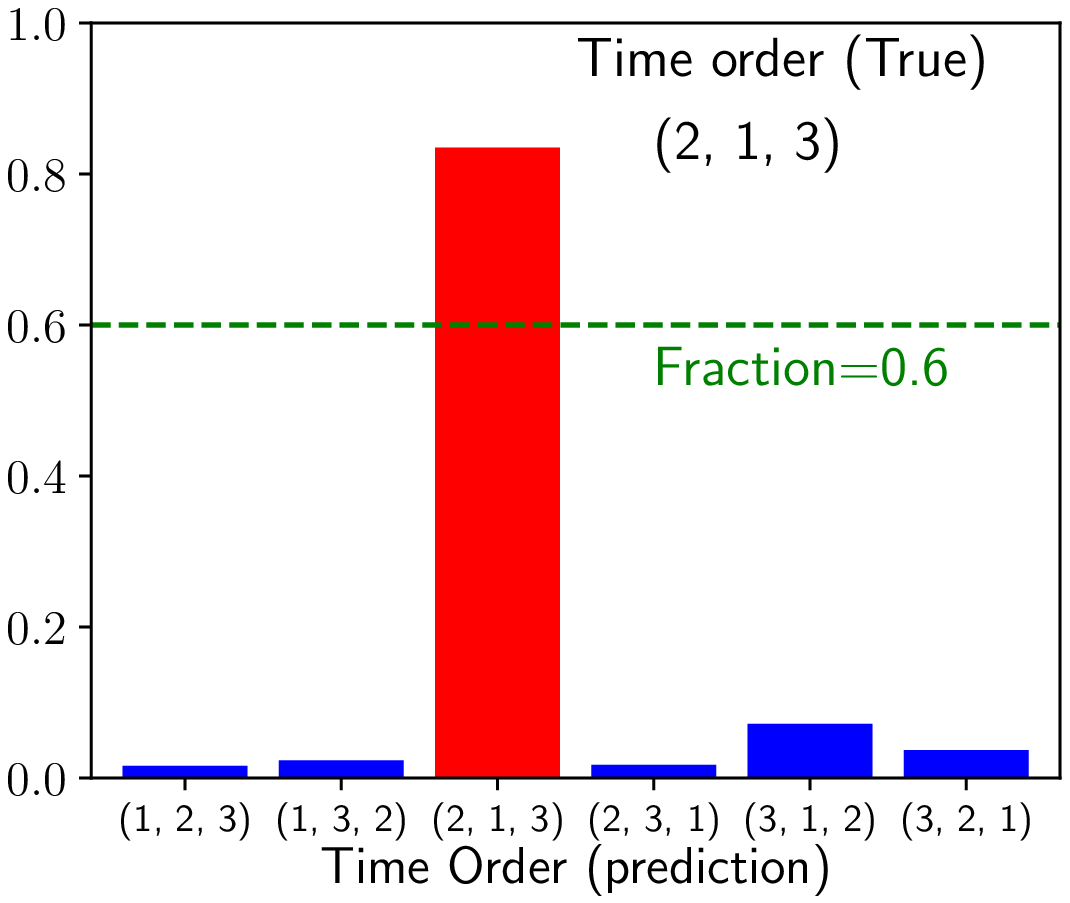}
        \end{minipage}\\
        \begin{minipage}{0.33\hsize}
            \centering
            \includegraphics[width=5.15cm]{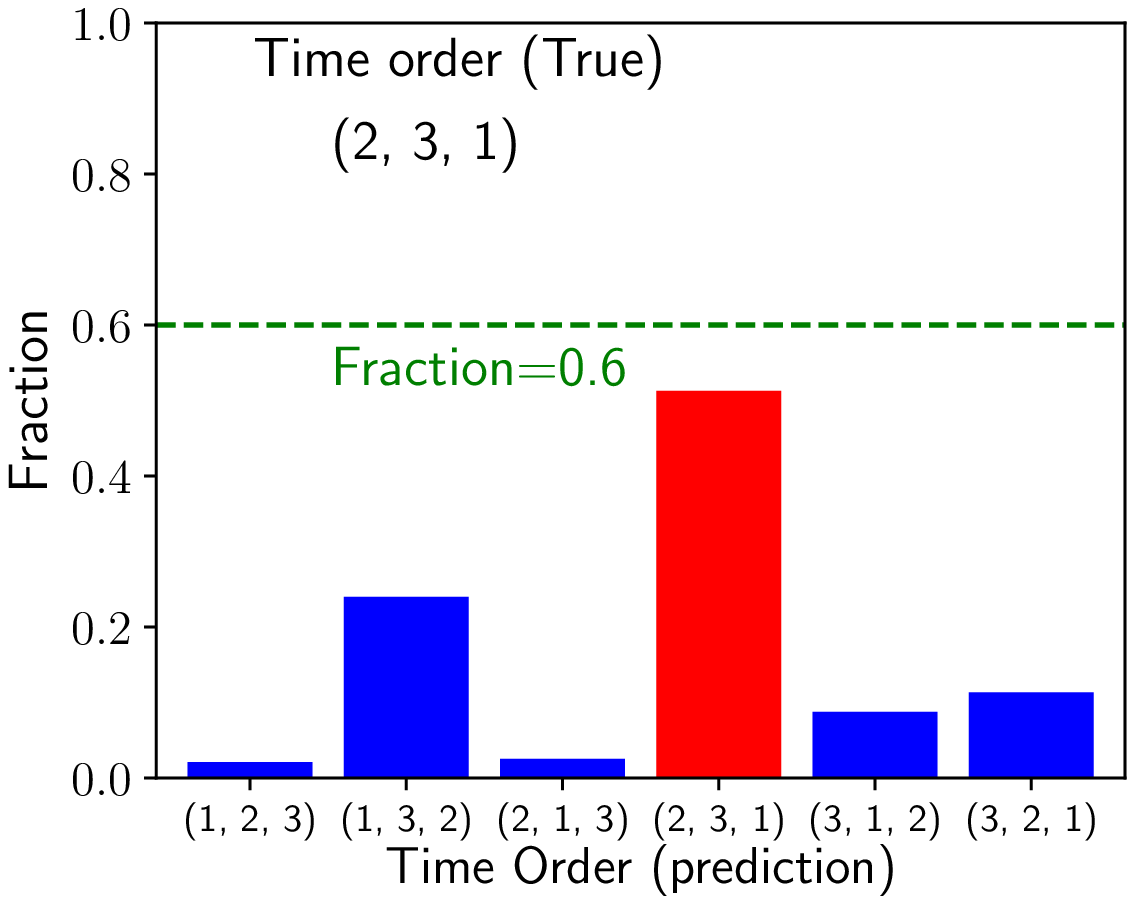}
        \end{minipage}
        \begin{minipage}{0.33\hsize}
            \centering
            \includegraphics[width=5.15cm]{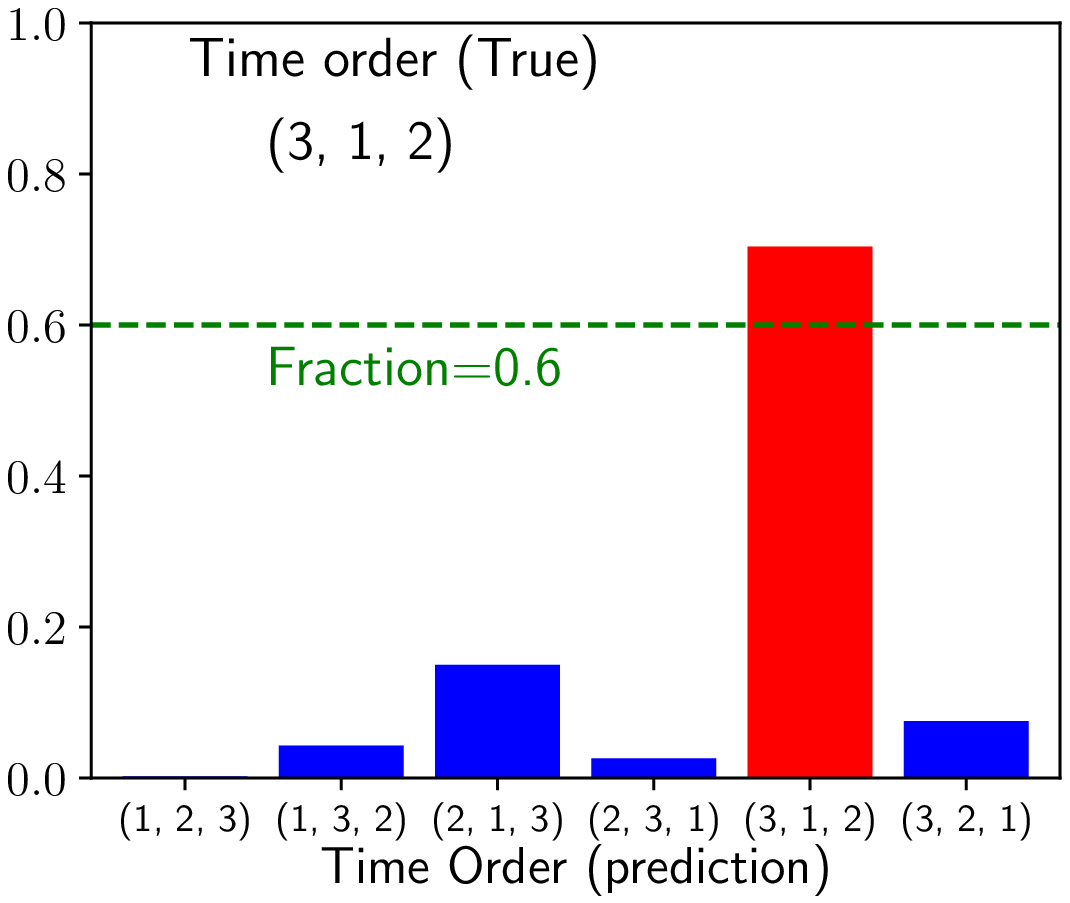}
        \end{minipage}
        \begin{minipage}{0.33\hsize}
            \centering
            \includegraphics[width=5.15cm]{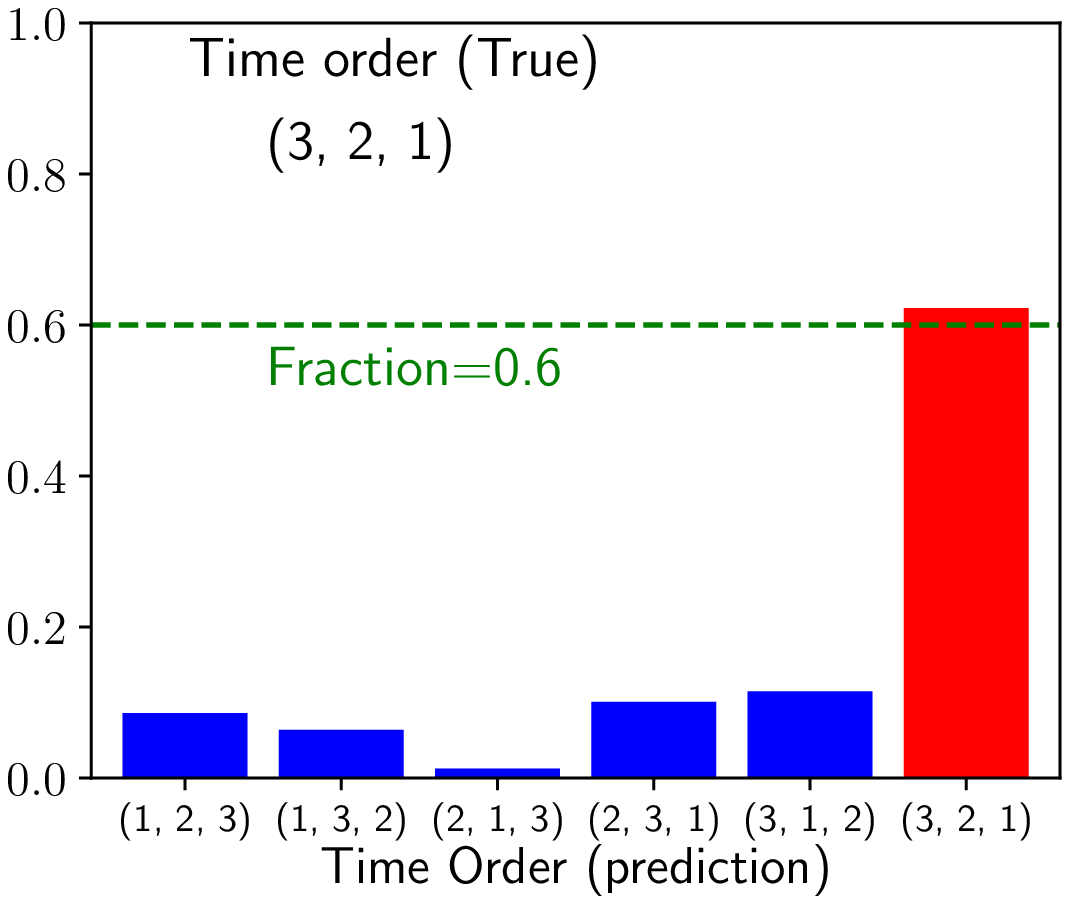}
        \end{minipage}
    \end{tabular}
    \caption{Fractions of predicted hit order labels by the neural network model for each true label in $1.4\,\mathrm{MeV}$ three-hit events. Red bars indicate fractions of correctly predicted events while blue bars are those of misclassified events. The fractions of properly predicted events are better than 60\% (dashed green lines).}
    \label{fig:hit_order_accuracy_3hits}
\end{figure*}

Figure \ref{fig:hit_order_accuracy_3hits} shows hit order labels predicted by the network model for each true one in the case of three-hit events. Red and blue bars denote the correct and incorrect order labels, respectively. For almost all the true order labels, the accuracy of the model prediction is better than 60\% (the green dashed lines). This excellent performance indicates that the network training by the simulation data works effectively.

\subsection{Design of the network}
\label{sec:result_2}
It is important to optimize the configuration of the neural network because it has an enormous number of possible configurations of layers.
One of the most important factors is the number of the hidden layers.
In general, a network with many layers is flexible. However, the number should be small as long as the performance is good enough to avoid vanishing gradient and time-consuming calculation\cite{Bengio1994,Glorot2010}.
In addition to the three-layer network described above, we study five different networks, as listed in Table \ref{tab:multi_layer}, to evaluate how the complexity of a network improves the performance.
\begin{table*}
    \centering
    \begin{threeparttable}[h]
        \caption{Ratios $r_{\pm20^\circ}$ of six different models for \SI{1.4}{MeV} photon events}
        \label{tab:multi_layer}
        \begin{tabular}{rlrrrrrr}\hline
            Model & Configuration\tnote{a} & \multicolumn{6}{c}{$r_{\pm20^\circ}$ [\%]} \\
                 &                  & 3hits & 4hits & 5hits & 6hits & 7hits & 8hits\\ \hline\hline
        1 layer  & (64)             & 49.3  & 51.6  & 49.3  & 46.2  & 42.3  & 37.0\\
                 & (128)            & 53.5  & 53.0  & 57.0  & 46.9  & 56.9  & 63.1\\
                 & (256)            & 51.4  & 52.7  & 58.4  & 58.6  & 66.4  & 64.0\\
        2 layers & (256,128)        & 52.6  & 55.6  & 65.1  & 67.8  & 68.4  & 66.0\\
        3 layers & (256,128,64)     & 56.3  & 59.3  & 67.7  & 69.6  & 70.0  & 65.7\\
        4 layers & (256,256,128,64) & 57.4  & 62.0  & 68.7  & 71.6  & 70.9  & 67.1\\ \hline
        \end{tabular}
        \begin{tablenotes}
            \item[a] Node number in each hidden layer. 
        \end{tablenotes}
    \end{threeparttable}
\end{table*}

\begin{figure*}[ht]
    \centering
    \begin{tabular}{c}
        \begin{minipage}{0.33\hsize}
            \centering
            \includegraphics[width=5.0cm]{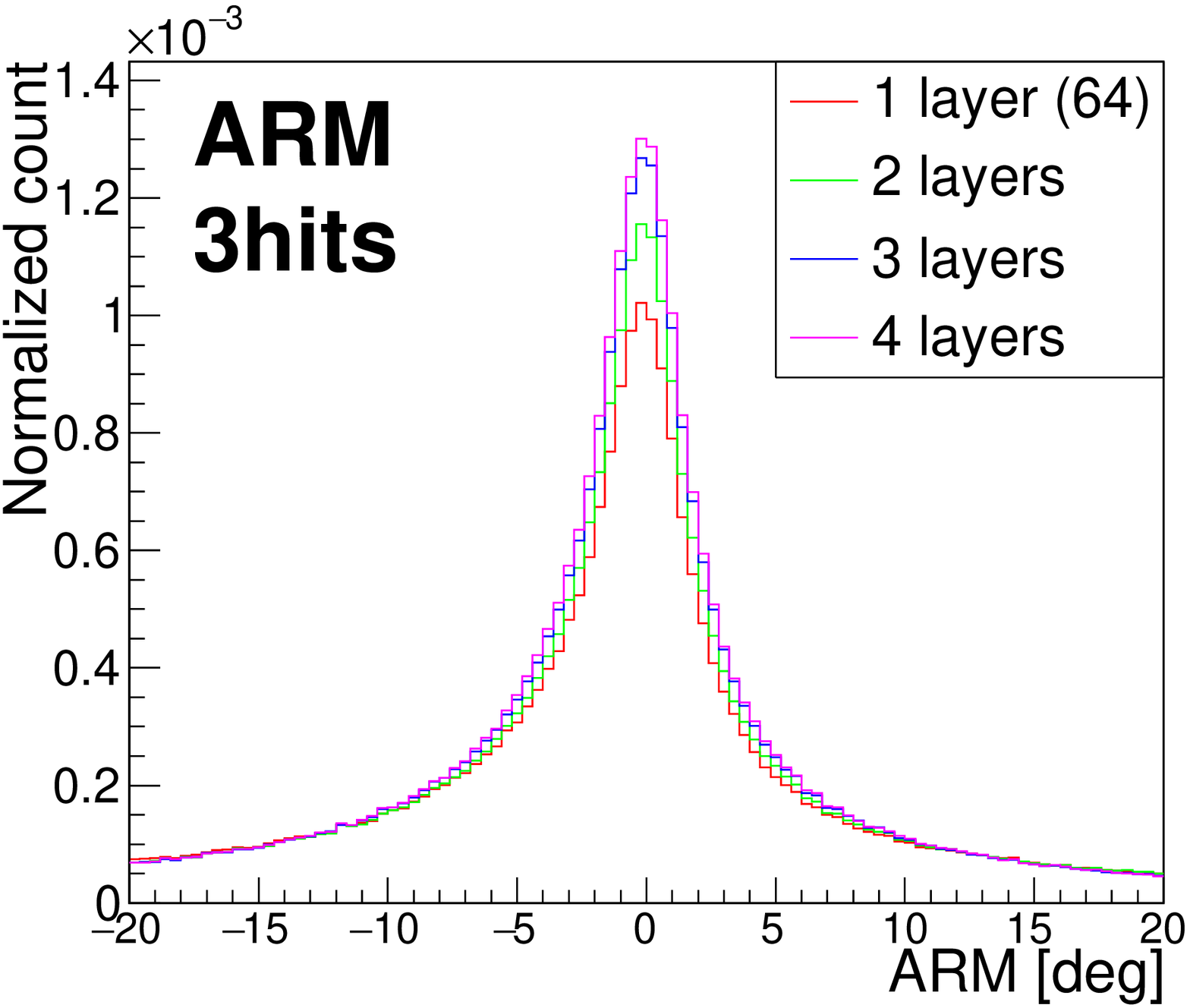}
        \end{minipage}
        \begin{minipage}{0.33\hsize}
            \centering
            \includegraphics[width=5.0cm]{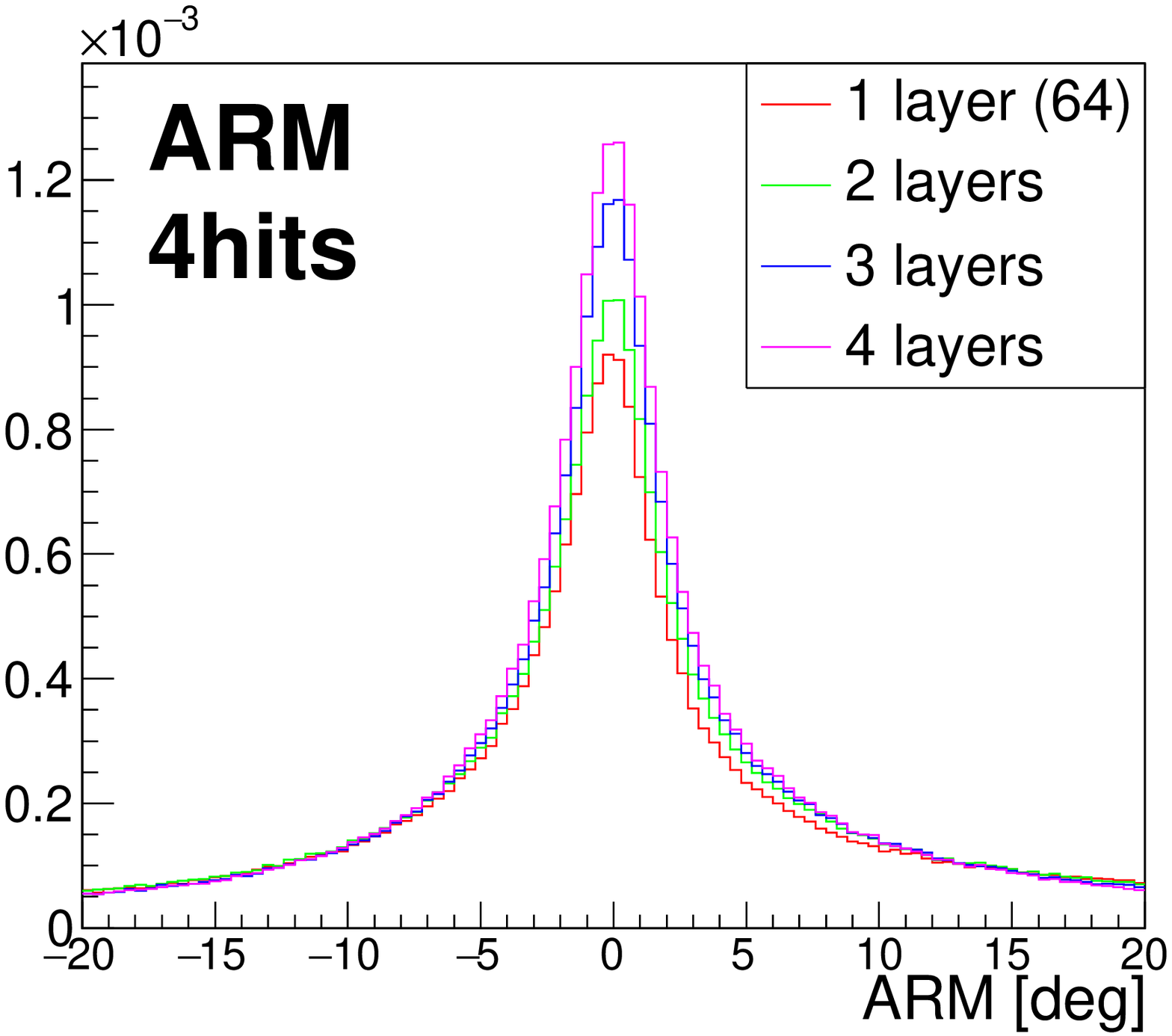}
        \end{minipage}
        \begin{minipage}{0.33\hsize}
            \centering
            \includegraphics[width=5.0cm]{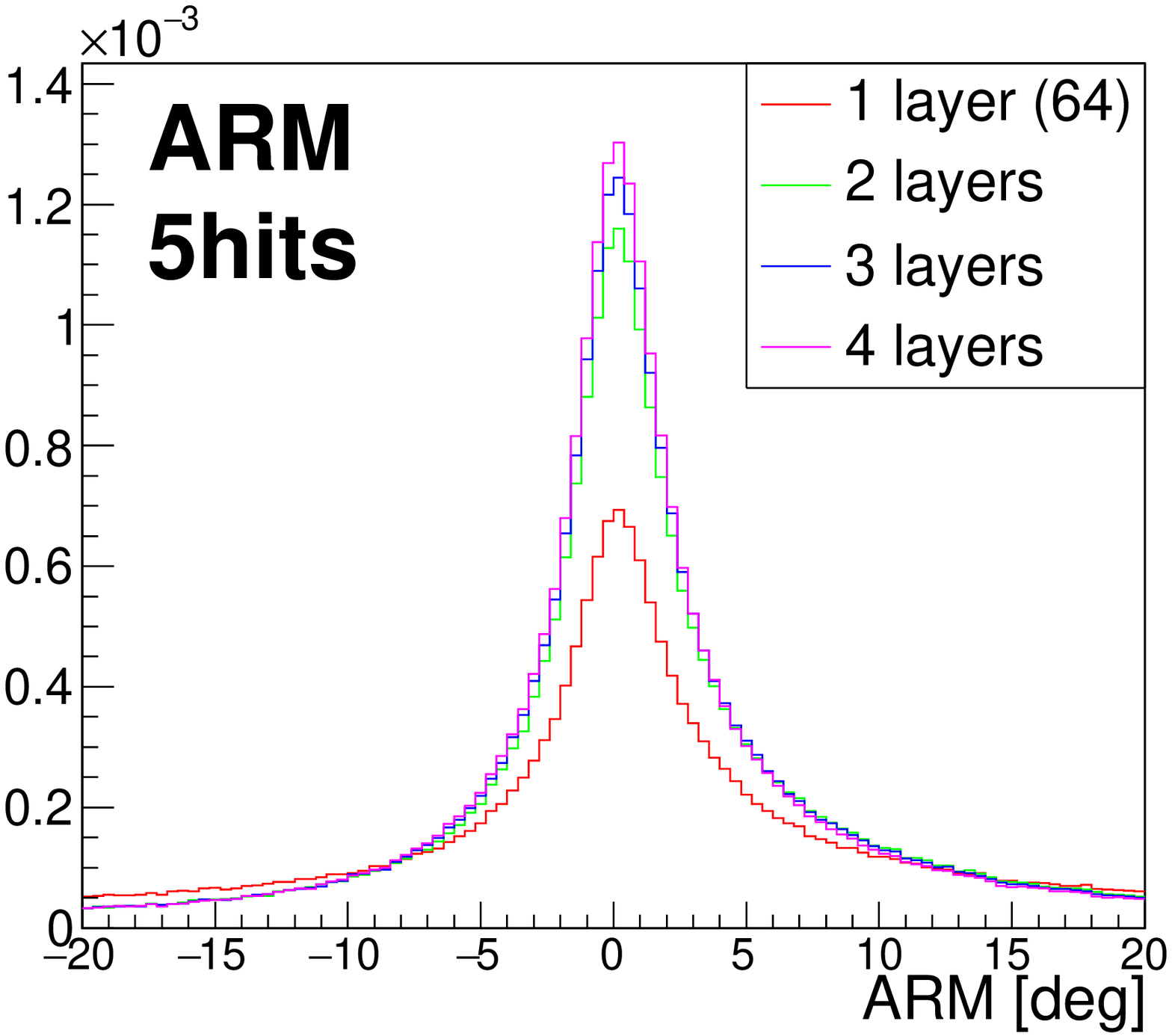}
        \end{minipage}\\
        \begin{minipage}{0.33\hsize}
            \centering
            \includegraphics[width=5.0cm]{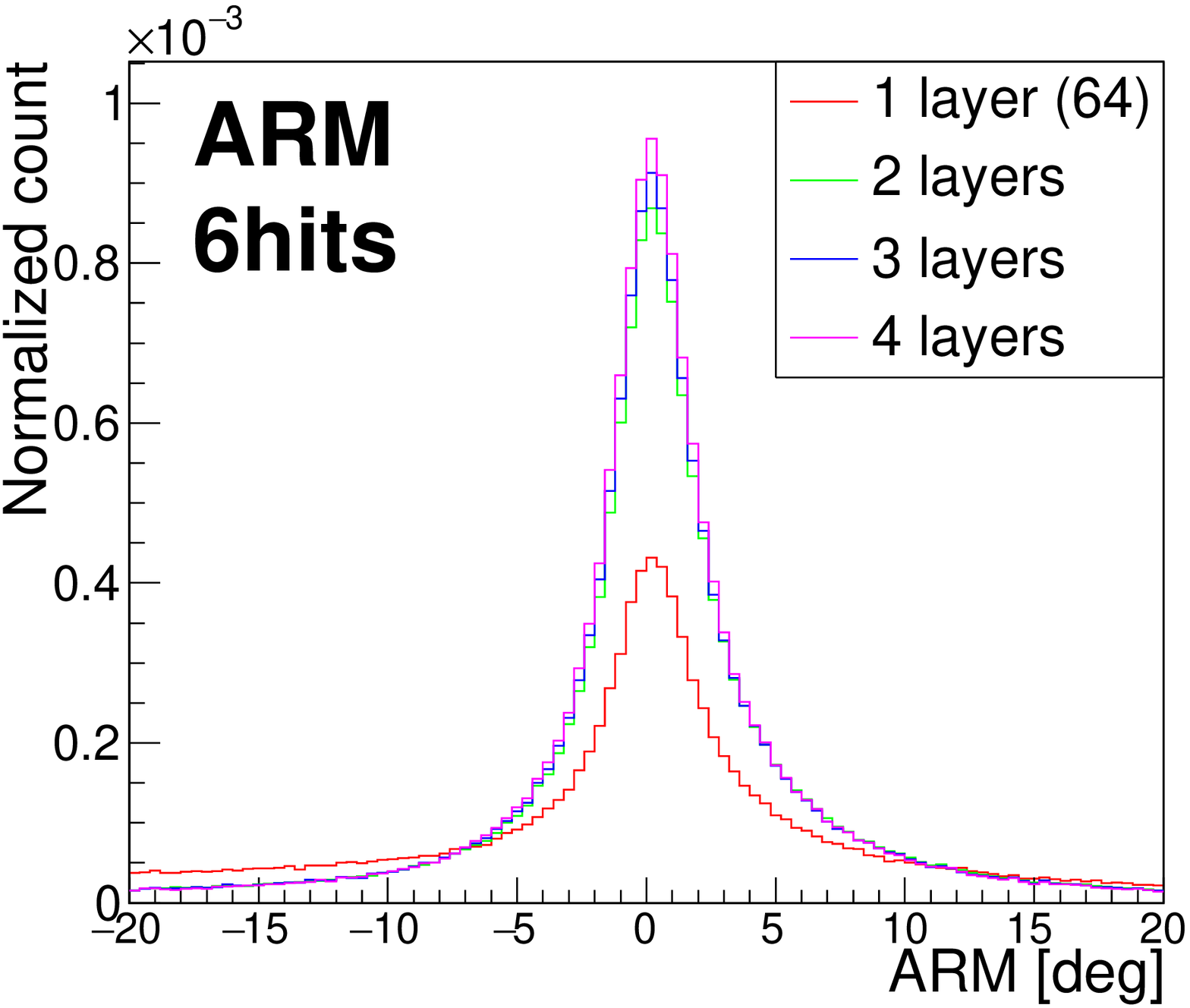}
        \end{minipage}
        \begin{minipage}{0.33\hsize}
            \centering
            \includegraphics[width=5.0cm]{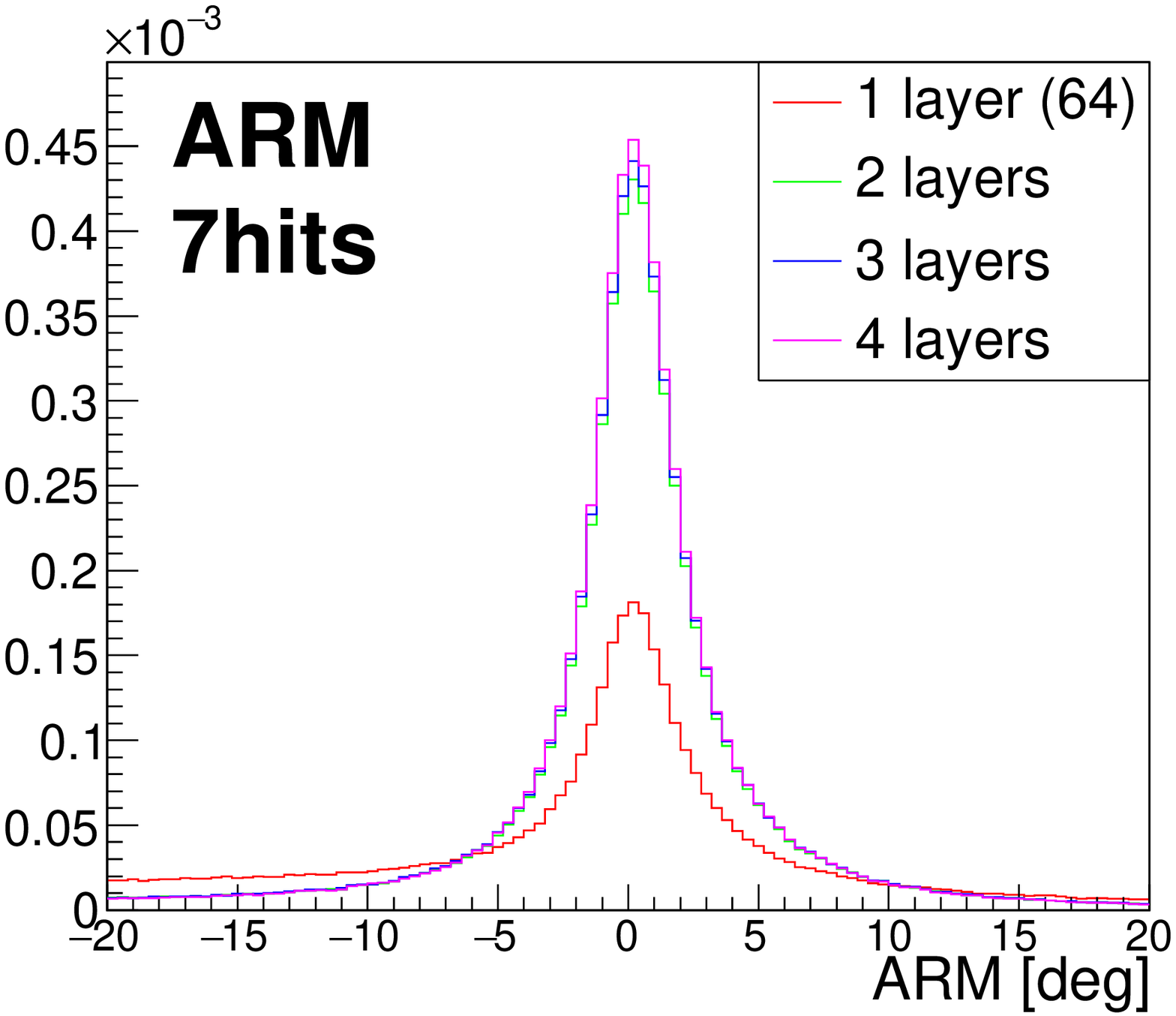}
        \end{minipage}
        \begin{minipage}{0.33\hsize}
            \centering
            \includegraphics[width=5.0cm]{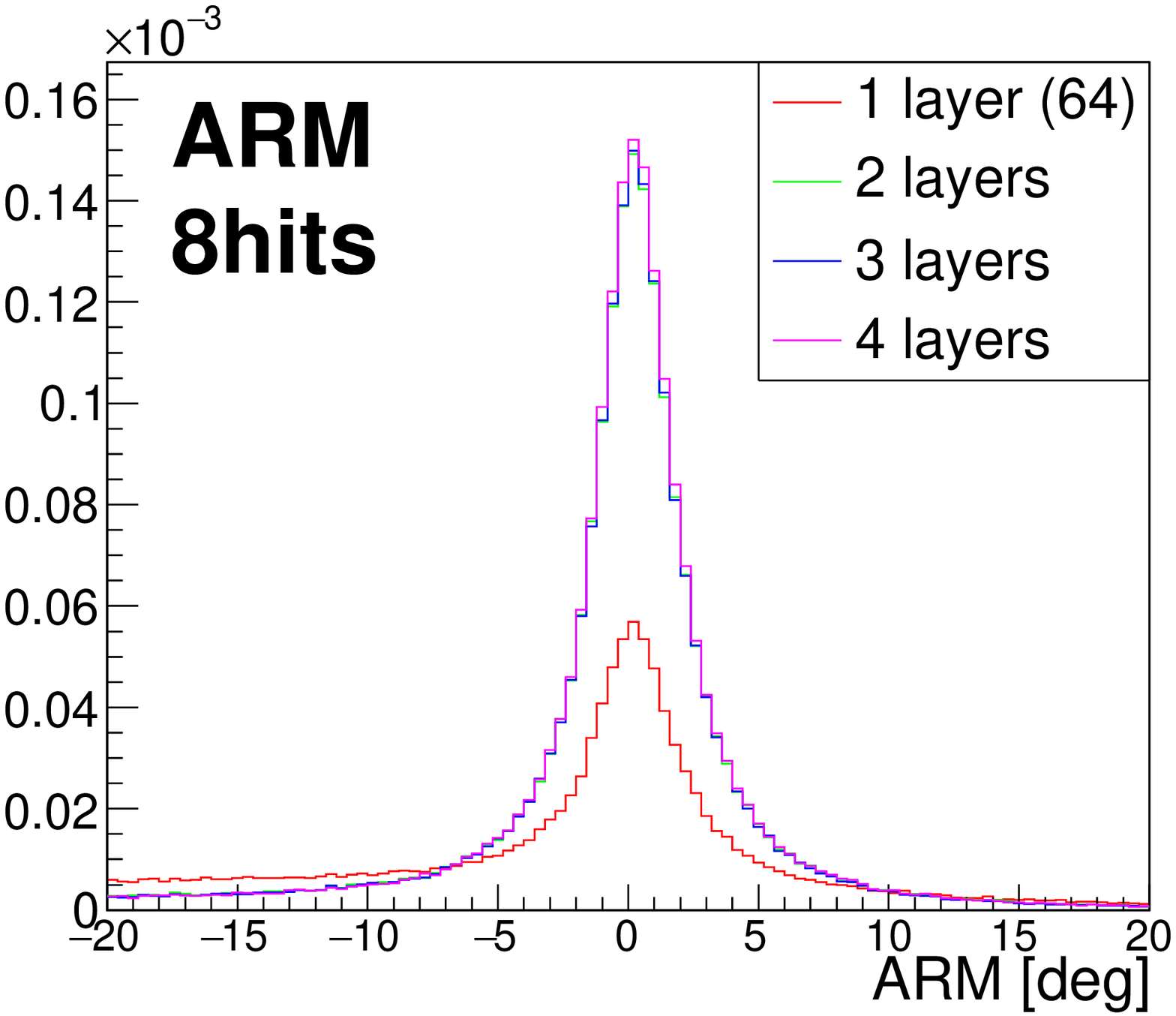}
        \end{minipage}
    \end{tabular}
    \caption{ARM distributions of reconstructed \SI{1.4}{MeV} events zoomed around $\mathrm{ARM}=0$ for different layer models. It can be seen that Peaks around \SI{1.4}{MeV} get sharper as the number of layers increases.}
    \label{fig:arm_energy_layers}
\end{figure*}

The performance of the event reconstruction is evaluated by a reconstruction efficiency $r_{\pm20^\circ}$, which is defined as a fraction of events whose ARMs are within a range of $[\ang{-20},\,\ang{20}]$ in all the reconstructed events.
Table \ref{tab:multi_layer} shows values of $r_{\pm20^\circ}$ for each hit order label and the five tested models in the data of incident \SI{1.4}{MeV} photons.
It is noticeable that $r_{\pm20^\circ}$ becomes more than \SI{50}{\%} for any hit order labels when we use two hidden layers.
Figure \ref{fig:arm_energy_layers} shows the distributions of the ARM reconstructed by four models that have different numbers of hidden layers.
The peaks around $\mathrm{ARM}=0$ get sharper as the numbers of layers increase.
While adding more hidden layers makes a model more flexible, it requires a longer time for training and analysis.
We choose the three-layer model in this work by considering the performance and the computational costs.

\subsection{Comparison with conventional reconstruction algorithm}
\label{sec:result_3}
We compare the present neural network method with other reconstruction algorithms; the classical $\chi^2$ model\cite{Oberlack2000} and the physics-based probabilistic model\cite{Yoneda2022}.
The classical $\chi^2$ model considers all the possible hit orders of $n!$ and chooses one sequence that minimizes an evaluation function $\chi^2$:
\begin{equation}
    \chi^2 = \sum_{i=2}^{N-1}\frac{\left(\cos{\theta_{\mathrm{K},i}}-\cos{\theta_{\mathrm{G},i}}\right)^2}{\sigma^2_{\cos{\theta_{\mathrm{K},i}}}+\sigma^2_{\cos{\theta_{\mathrm{G},i}}}}\ ,
\end{equation}
where $\theta_{\mathrm{K},i}$ denotes the scattering angle of $i$-th hit calculated kinematically (Equation (\ref{eq:compton})), $\theta_{\mathrm{G},i}$ denotes the same angle but calculated geometrically (Equation (\ref{eq:cos_geometry})), and $\sigma^2_{\cos{\theta_{\mathrm{K},i}}}$ and $\sigma^2_{\cos{\theta_{\mathrm{G},i}}}$ are the squared errors of $\cos{\theta_{\mathrm{K},i}}$ and $\cos{\theta_{\mathrm{G},i}}$, respectively. If reconstruction of the classical model is performed well, the value $\chi^2$ will be near 1. Therefore, incorrectly reconstructed events can be eliminated from the analysis by rejecting events whose $\chi^2$ is larger than a threshold prescribed beforehand. This selection, however, inevitably reduces efficiency.

For the physics-based probabilistic model, an escape flag is considered in addition to prediction of hit orders.
A hit order and an escape flag are determined by evaluating a probability function of each event type candidate and selecting one of them that yields the maximum function value.

\begin{figure}[htbp]
    \centering
    \includegraphics[width=6.5cm]{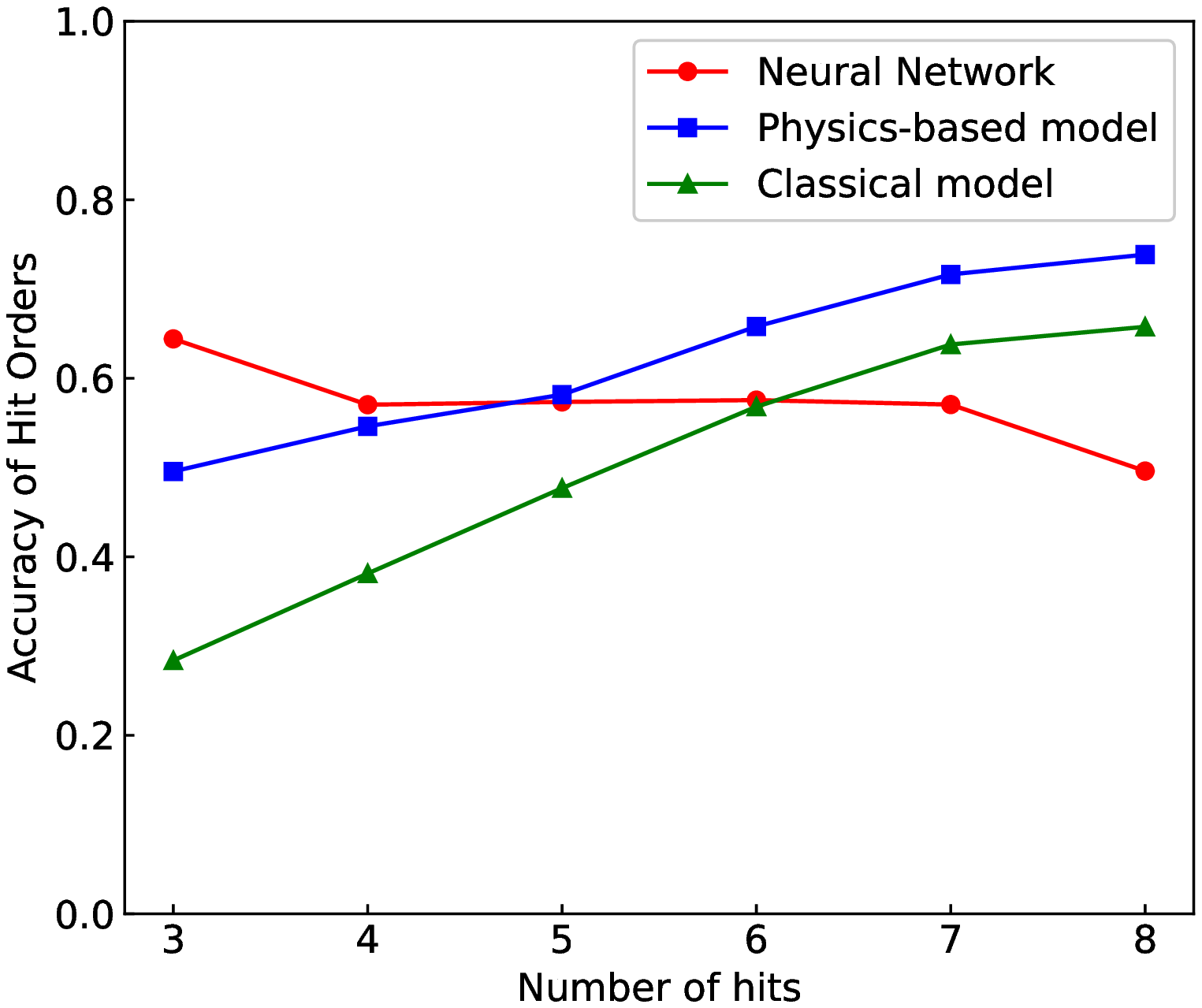}
    \caption{Accuracies of hit orders of $1.4\,\mathrm{MeV}$ monoenergetic photons. The accuracies of the neural network model (red points) are better performance in fewer scattering events compared with the physics-based one (blue boxes) and the classical one (green triangles).}
    \label{fig:accuracy_hitorder}
\end{figure}

\begin{figure}[htbp]
    \centering
    \includegraphics[width=6.5cm]{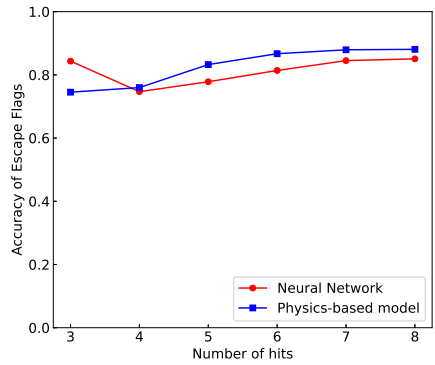}
    \caption{Accuracies of escape flags of $1.4\,\mathrm{MeV}$ monoenergetic photons.}
    \label{fig:accuracy_escflag}
\end{figure}

Figure \ref{fig:accuracy_hitorder} shows the accuracies of hit order labels of the neural network model (red points), the physics-based probabilistic one (blue boxes), and the classical one (green triangles).
For all the methods, we use the same simulation data of \SI{1.4}{MeV} monoenergetic photons in Section 4.1.
Generally, the neural network model works well, particularly when the number of hits is small, 3 or 4, for which escape events are influential.
Figure \ref{fig:accuracy_escflag} demonstrates that the performance of the neural network model to discriminate between fully-absorbed and escape events is almost equal to that of the physics-based one.

\section{Discussion}
\label{sec:discussion}
We have constructed a new method to deal with multiple Compton scattering events by a neural network.
While most other approaches assume that events are fully absorbed, our multi-task neural network model predicts the scattering order and the escape flag in the same framework.
This improves detection efficiencies of next-generation Compton telescopes.
In this section, we discuss the advantages unique to neural networks and possible improvements of our model.

This paper does not deal with two-hit events since a fraction of those events is a minority, less than $\sim 28$\% of $n$-hit $(n \geq 2)$ events, in an energy range of 0.5--\SI{3}{MeV}, assuming the liquid argon detector described in Section \ref{sec:numerical_experiments_1}.
The GRAMS project focuses on a relatively high energy band, and therefore an important motivation of this work is to develop an event reconstruction algorithm that treats multiple scattering events, which constitute the majority of the Compton scattering events above \SI{1}{MeV}.
In a low-energy part of the sub-MeV band, below \SI{0.3}{MeV}, the two-hit events become dominant and should be considered for an efficient observation.

Event reconstruction of real detector data has many complicated factors to be considered, unlike a simplified detector simulation such as non-uniformity of a detector response including a dead or inactive volume and inappropriately merged hits of multiple different photon interactions.
These effects are difficult to consider in analytical physics models\cite{Yoneda2022}.
In contrast, neural networks can construct almost arbitrary maps if sufficient hidden layers are stacked, and such a complicated detector response is automatically considered in the training data. In this data-driven way, the neural network model can be applied to other types of detectors that adopt semiconductor and gas detectors, while a liquid argon time projection chamber is assumed in this paper.

In general, the performance of a neural network model can improve by adding other input variables (see e.g. \cite{Zoglauer2007a}).
In this present model, we input pairs of energy deposits and position coordinates as essential information on the physical interactions within the Compton telescope.
Because other critical input parameters depend on problems and detector setup, specific optimization is needed for each system. This will be considered for more practical detector configuration in the future.

When a Compton telescope is transported to a balloon altitude or a satellite orbit, many background particles---cosmic rays, geomagnetically trapped particles, particles generated by interactions of cosmic rays with the atmosphere, and so on---are expected to enter the detector.
Our network model has the possibility to deal with them without a large-scale modification.
For this purpose, we only have to add an event type classification such as a particle type or a signal-or-background flag to the network we already developed. 

In this paper, we perform the numerical experiment that considers an energy band up to \SI{3.0}{MeV}, since there are many nuclear lines interesting in astrophysical aspects in this band. On the other hand, some lines higher than \SI{3.0}{MeV} are also important, \SI{4.4}{MeV} ($^{12}\mathrm{C}^{\ast}$ line), \SI{6.13}{MeV} ($^{16}\mathrm{O}^{\ast}$ line). However, pair creation cannot be ignored in such a high energy band, and which makes the energy reconstruction difficult. A possible solution is to introduce a function that predicts whether a photon results in pair creation or not (a pair creation flag) like a signal-or-background flag as already described.

\section{Conclusions}
\label{sec:conclusion}
We developed a multi-task neural network model for reconstructing multiple scattering events, including escape events.
Our model can be applied to a wide variety of Compton telescopes since this model utilizes only basic outputs of a gamma-ray detector, including the energy deposits and positions of gamma-ray interactions.
To validate the model, we conducted numerical experiments using Monte Carlo simulation, assuming a large-area Compton telescope using liquid argon to measure isotropic $4\pi$ gamma rays up to \SI{3.0}{MeV}.
The reconstruction model showed excellent performances of the event reconstruction and the computation costs for multiple scattering events that consist of up to eight hits.
The accuracy of hit order prediction was around \SI{60}{\%} while that of escape flags was higher than \SI{70}{\%} for up to eight-hit events as shown in Figure \ref{fig:accuracy_hitorder} and Figure \ref{fig:accuracy_escflag}.
In this simplified test, the present neural network method demonstrates its better performance in terms of energy and angular resolutions, especially for escape events.
Since the network model is data-driven and automatically optimized by simulation data, it can be flexible and robust against more complicated detector responses of real instruments.

\section{Acknowledgement}
\label{sec:acknowledgement}
The authors acknowledge all the GRAMS project members who have contributed to an essential discussion on this study.
The authors also thank anonymous reviewers for their useful comments which improve the manuscript.
This work has been supported by KAKENHI Grant-in-Aid for Scientific Research on Innovative Areas No.\ 18H05458, 18H05861, 19H01906, 19K03908, 19H05185,  20K14524, 20K20527, 20K22355, and 22H00128, RIKEN Incentive Research Projects, and by Toray Science and Technology Grant No.\ 20-6104 (Toray Science Foundation).
ST is supported by RIKEN Junior Research Associate Program, and JSPS Research Fellowship for Young Scientist (No.\ 22J11653).
Y.I. is supported by JSPS KAKENHI Grant Number JP18H05458 and JP19K14772.


 \bibliographystyle{elsarticle-num} 
 \bibliography{cas-refs}





\end{document}